\pdfoutput=1
\UseRawInputEncoding

\documentclass{egpubl}

\makeatletter
\copyrightTextTitPag{}%
\copyrightTextRunPag{}%
\makeatother

\makeatletter
\def\ps@titlepage{\let\@mkboth\@gobbletwo
  \def\@oddhead{}\def\@evenhead{}%
  \def\@oddfoot{}\def\@evenfoot{}%
}
\makeatother

\makeatletter
\def\ps@headings{\let\@mkboth\markboth
  \def\@oddhead{}\def\@evenhead{}%
  \def\@oddfoot{\hfil\thepage\hfil}%
  \def\@evenfoot{\hfil\thepage\hfil}%
  \def\sectionmark##1{}%
  \def\subsectionmark##1{}%
}
\makeatother

\AtBeginDocument{%
  \hypersetup{
    pdfsubject={},
    pdfkeywords={preprint, OSF},
  }%
}
\usepackage[T1]{fontenc}
\usepackage{dfadobe}

\usepackage{cite}  
\BibtexOrBiblatex
\electronicVersion
\PrintedOrElectronic
\ifpdf \usepackage[pdftex]{graphicx} \pdfcompresslevel=9
\else \usepackage[dvips]{graphicx} \fi

\usepackage{egweblnk}

\usepackage{lipsum}
\usepackage{graphicx}
\usepackage{booktabs}
\usepackage{soul}

\usepackage{array}
\newcolumntype{L}[1]{>{\raggedright\arraybackslash}p{#1}}

\usepackage{array}
\usepackage{enumitem} 
\usepackage{enumitem} 
\usepackage{multirow} 

\usepackage{makecell}

\usepackage{xcolor}
\definecolor{assorange}{HTML}{f9cb9c}
\definecolor{ontpurple}{HTML}{b4a7d6}
\definecolor{mechgreen}{HTML}{b6d7a8}
\definecolor{popyellow}{HTML}{ffe599}
\definecolor{intblue}{HTML}{a4c2f4}

\newcommand{\asshl}[1]{\sethlcolor{assorange}\hl{#1}}
\newcommand{\onthl}[1]{\sethlcolor{ontpurple}\hl{#1}}
\newcommand{\mechhl}[1]{\sethlcolor{mechgreen}\hl{#1}}
\newcommand{\pophl}[1]{\sethlcolor{popyellow}\hl{#1}}
\newcommand{\inthl}[1]{\sethlcolor{intblue}\hl{#1}}

\usepackage{tcolorbox}
\tcbuselibrary{listings,breakable}
\tcbset{
  assbox/.style={
    colback=gray!5, 
    colframe=assorange, 
    fonttitle=\bfseries, 
    coltitle=black, 
    boxrule=1pt, 
    left=1mm, 
    right=1mm, 
    top=1mm, 
    bottom=1mm,
    parskip=1em
  }
}
\tcbset{
  ontbox/.style={
    breakable,
    colback=gray!5, 
    colframe=ontpurple, 
    fonttitle=\bfseries, 
    coltitle=black, 
    boxrule=1pt, 
    left=1mm, 
    right=1mm, 
    top=1mm, 
    bottom=1mm,
    parskip=1em
  }
}
\tcbset{
  mechbox/.style={
    colback=gray!5, 
    colframe=mechgreen, 
    fonttitle=\bfseries, 
    coltitle=black, 
    boxrule=1pt, 
    left=1mm, 
    right=1mm, 
    top=1mm, 
    bottom=1mm,
    parskip=1em
  }
}
\tcbset{
  popbox/.style={
    colback=gray!5, 
    colframe=popyellow, 
    fonttitle=\bfseries, 
    coltitle=black, 
    boxrule=1pt, 
    left=1mm, 
    right=1mm, 
    top=1mm, 
    bottom=1mm,
    parskip=1em
  }
}
\tcbset{
  intbox/.style={
    colback=gray!5, 
    colframe=intblue, 
    fonttitle=\bfseries, 
    coltitle=black, 
    boxrule=1pt, 
    left=1mm, 
    right=1mm, 
    top=1mm, 
    bottom=1mm,
    parskip=1em
  }
}

\title[The State of the Art in Visualization Literacy]%
      {The State of the Art in Visualization Literacy}

\author[M. Varona, K. Bonilla, M. Hedayati, A. Joshi, M. Kay, L. Harrison, C. Nobre]
{\parbox{\textwidth}{\centering Matthew Varona$^{1}$, Karen Bonilla$^{2}$, Maryam Hedayati $^{3}$, Alark Joshi$^{4}$, Matthew Kay$^{3}$, Lane Harrison$^{2}$, and Carolina Nobre$^{1}$ 
        }
        \\
{\parbox{\textwidth}{\centering $^1$University of Toronto, Canada\\
       $^2$ Worcester Polytechnic Institute, United States\\
       $^3$ Northwestern University, United States \\
       $^4$ University of San Francisco, United States
       }
}
}
\begin{document}

\teaser{
 \includegraphics[width=1\linewidth]{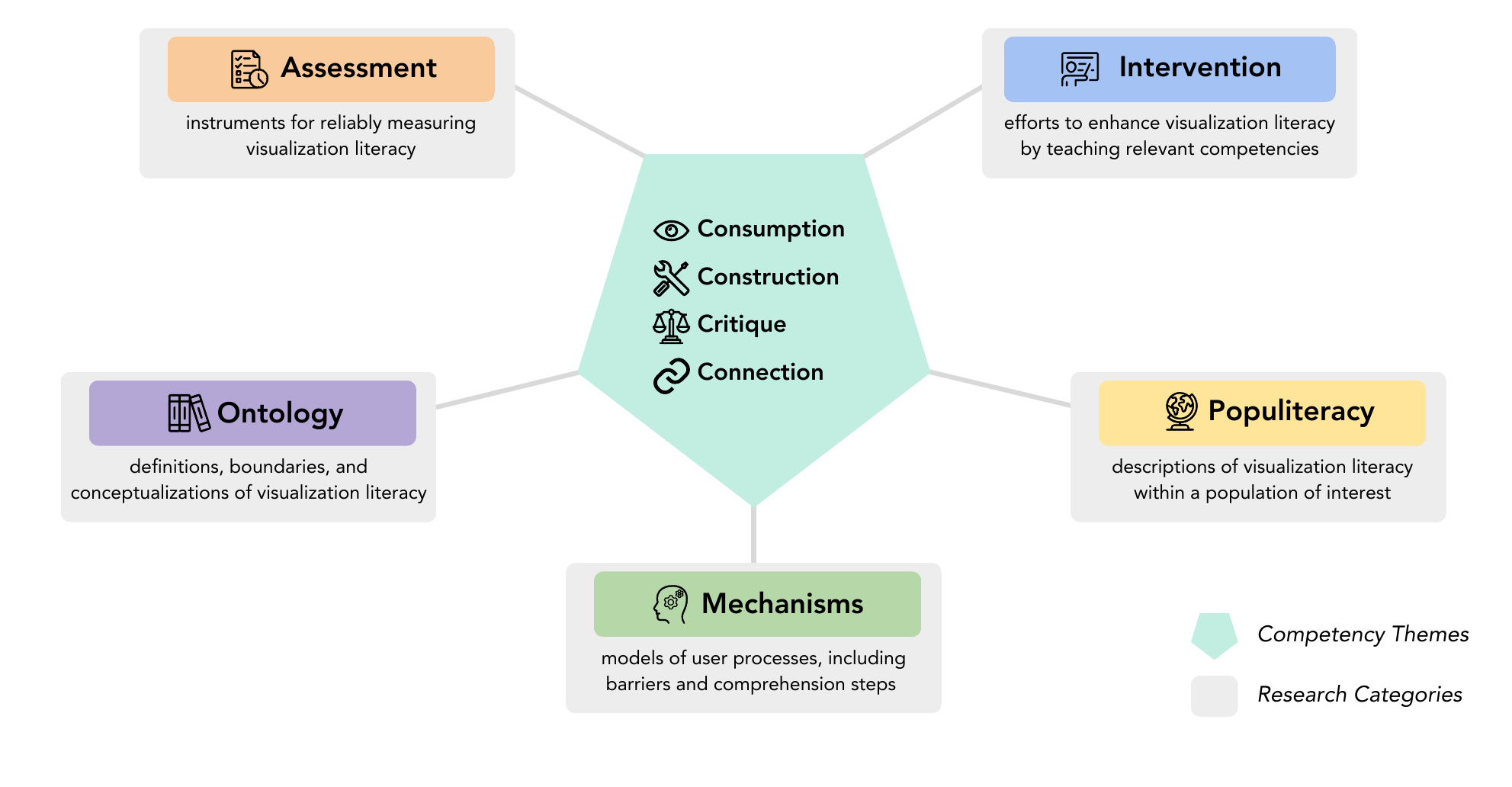}
 \centering
  \caption{A taxonomy of visualization literacy uniting \textbf{research categories} and \textbf{competency themes}. Research categories provide structure to the types of contributions found in our survey corpus of 374 papers. Competency themes broadly describe the skills necessary to engage with visualizations.}
\label{fig:teaser}
}

\maketitle
\begin{abstract}
Research in visualization literacy explores the skills required to engage with visualizations. This state-of-the-art report surveys the current literature in visualization literacy to provide a comprehensive overview of the field. We propose a taxonomy of visualization literacy that organizes the field into competency themes and research categories. To address ambiguity surrounding the term ``visualization literacy'', we provide a framework for operationalizing visualization literacy based on application contexts (including domain, scenario, and audience) and relevant competencies, which are categorized under consumption, construction, critique, and connection. Research contributions are organized into five categories: ontology, assessment, mechanisms, populiteracy, and intervention. For each category, we identify key trends, discuss which competencies are addressed, highlight open challenges, and examine how advancements within these areas inform and reinforce each other, driving progress in the field.

\end{abstract}  

\section{Introduction}

While literacy originally referred to the ability to read and write, the metaphor of ``literacy'' is now often used to describe various groups of related competencies, such as health literacy, media literacy, and information literacy. Among the fields that have adopted the term, it is perhaps visualization literacy that most closely matches early conceptualizations of literacy as reading and writing. Analogous to a shared set of symbols (letters) from which meaning is derived, visualization literacy refers to the ability to meaningfully engage with visual representations of data. 

Visualization literacy has received increased attention from the visualization research community over the past few years. Driven by the increased ubiquity of visualizations in daily life, researchers have sought to define, model, and measure visualization literacy, investigate factors that influence visualization comprehension, and improve visualization literacy through various interventions \cite{ChevalieretalObservationsReflectionsVisualization2018, LeeKimKwonVLATDevelopmentVisualization2017, NobreetalReadingPixelsInvestigating2024, CambaCompanyByrdIdentifyingDeceptionCritical2022,PengetalEvaluatingBloomsTaxonomybased2022}. This increase in research appetite is further demonstrated through workshops at IEEE VIS and ACM CHI \cite{KecketalEduVisWorkshopVisualization2023, GeetalMoreComprehensiveUnderstanding2024}, the recent Dagstuhl Seminar \cite{BachetalVisualizationEmpowermentHow2023}, and a special issue in IEEE Computer Graphics and Applications \cite{BachetalSpecialIssueVisualization2021}, which have collectively sought to advance our understanding of visualization literacy and encourage discourse about the topic. 

Visualization literacy has been studied not only by visualization researchers, but in related disciplines where visual communication of data plays an important role. For example, health risk communication research has investigated the role of individual differences in visualization reading ability on the efficacy of visual aids \cite{Garcia-RetameroCokelyDesigningVisualAids2017, Garcia-RetameroCokelyCommunicatingHealthRisks2013}. The ability to confidently use visualizations has also been identified by science educators as a gateway for building disciplinary expertise \cite{DanieletalDefinitionRepresentationalCompetence2018}. These varied contexts of study have given rise to different terms referring to similar phenomena, such as \textit{graph literacy} \cite{GalesicGarcia-RetameroGraphLiteracyCrossCultural2011}, \textit{graphicacy} \cite{BalchinGraphicacy1976}, and \textit{representational competence} \cite{DanieletalDefinitionRepresentationalCompetence2018}. While this survey includes research using a variety of related terms, we use \textit{visualization literacy} for consistency throughout the report. 

Accompanying the rise of visualization literacy research are challenges in scoping the field. The use of the literacy metaphor invites varied interpretations, resulting in differences in how visualization literacy is conceptualized, measured, and taught \cite{GeetalMoreComprehensiveUnderstanding2024}. For instance, while visualization literacy is often described as encompassing both reading and creating visualizations \cite{BornerBueckleGindaDataVisualizationLiteracy2019, SolenScopingFutureVisualization2022, GeetalMoreComprehensiveUnderstanding2024}, assessments in the field have predominantly focused on the reading of visualizations. This narrow focus overlooks other important competencies, such as designing effective visualizations or relating visualizations to the broader contexts in which they are used. Just as the concept of “literacy” has expanded beyond basic reading and writing to be viewed as a gateway to achieving one's goals, we argue that visualization literacy should be similarly broadened to reflect a fuller range of abilities and purposes. At the same time, clarifying the concept through more precise \textit{operationalization}---that is, breaking down an abstract idea into well-defined, researchable components---can help resolve some of the current ambiguities and support more coherent development in the field \cite{AdcockCollierMeasurementValidityShared2001}.

To address these challenges, our State-of-The-Art Report (STAR) synthesizes existing research on visualization literacy and identifies key opportunities for future work. Our contributions are as follows:
\begin{itemize}
    \item We organize existing work into a taxonomy of visualization literacy, providing an overview of concepts and techniques across five categories of research. (Section \ref{sect:taxonomy})
    \item We provide a framework of relevant competencies, derived from the analysis of prior work, that enables more precise operationalizations of visualization literacy. (Section \ref{fig:competencies})
    \item For each category, we summarize unsolved questions, highlight research implications for other categories, and explore opportunities for future work.
\end{itemize}

Our systematic review of visualization literacy research reveals several key insights about the field's current state and future directions. The field has made substantial progress in developing assessments of visualization skills and creating effective interventions for teaching visualization literacy. At the same time, our analysis reveals opportunities for growth - while some aspects of visualization literacy, such as the ability to read a visualization, are well-studied, others, like the ability to critique, receive less attention. Finally, while research has successfully characterized visualization literacy in several specific populations and contexts, significant gaps remain, particularly regarding non-Western contexts and accessibility considerations. These findings suggest important opportunities for expanding both the theoretical framework and practical applications of visualization literacy research.


\section{Related Surveys}
\label{sect:relatedsurveys}

This STAR is meant to provide a comprehensive and current overview of visualization literacy research. We found two prior surveys related to visualization literacy \cite{FiratJoshiLarameeInteractiveVisualizationLiteracy2022, SolenScopingFutureVisualization2022}, as well as a set of related papers that provide partial overviews of the field  \cite{BornerBueckleGindaDataVisualizationLiteracy2019, HedayatiHuntKayPixelsPracticesReconceptualizing2024, ChevalieretalObservationsReflectionsVisualization2018}. While they tend to be narrower in scope than our STAR, they provide valuable insight into specific aspects of visualization literacy.

Firat et al. \cite{FiratJoshiLarameeInteractiveVisualizationLiteracy2022} conducted a survey of 34 papers that studied users' visualization literacy. With a focus on evaluative approaches, they categorized papers based on different types of evaluations (crowdsourcing, classroom-based, controlled user study, or ``in the wild''). Similarly, Solen's survey of visualization literacy \cite{SolenScopingFutureVisualization2022} analyzed 20 papers and proposed three research themes, namely interpretation, construction, and believability. 

In order to address gaps in existing surveys, this STAR 1) employs a wider scope by reviewing 4,519 papers (374 in the final set), 2) encompasses a broader range of visualization literacy research, including papers from outside ``traditional'' visualization venues, 3) examines and categorizes diverse types of contributions in visualization literacy, including non-evaluative research such as interventions or theoretical models, and 4) provides a deeper exploration of visualization competencies and how they cut across various categories of research.

\section{Methodology}
\label{sect:methodology}
\begin{figure*}[htbp]
\centering
\includegraphics[width=\textwidth]{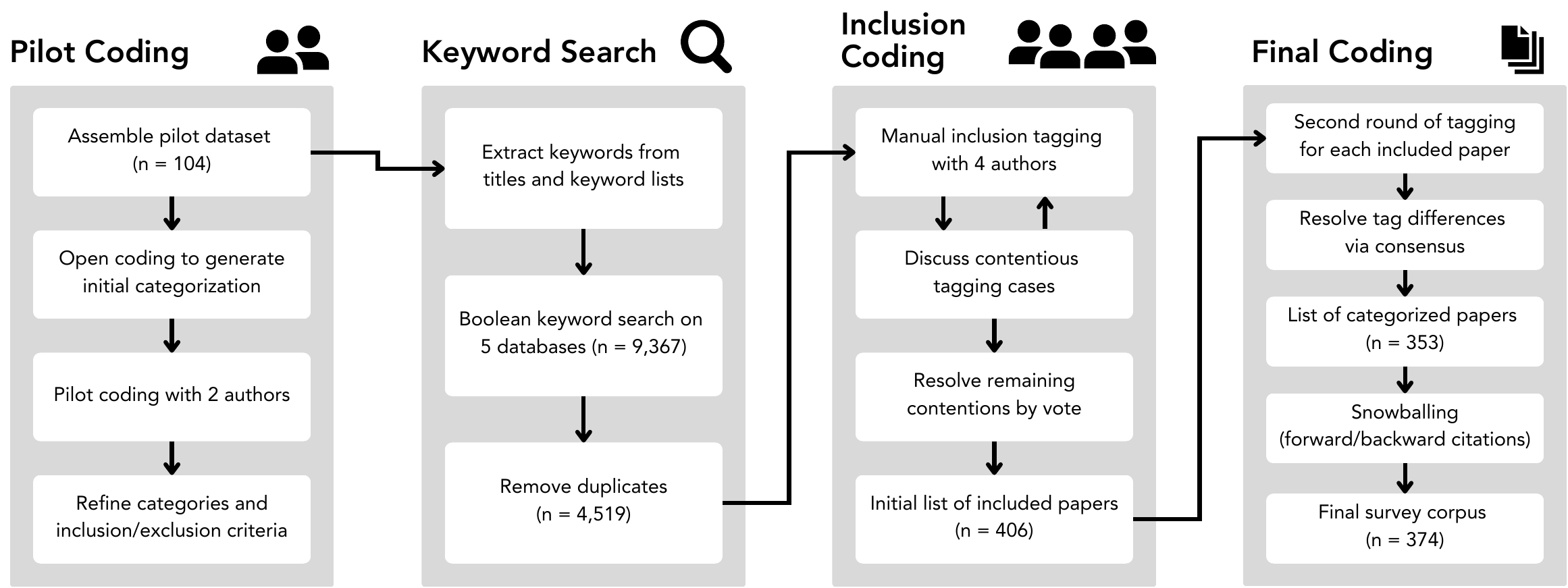}
\caption{The 4-stage methodology followed to acquire and tag relevant papers. In the \textbf{pilot coding} stage, an initial dataset of papers was coded to determine inclusion/exclusion criteria. \textbf{Keyword search} was conducted using phrases extracted from the pilot corpus. During the \textbf{inclusion coding} stage, the search results were each manually tagged with categories or marked for exclusion. Finally, each included paper was reviewed a second time during the \textbf{final coding} stage.}
\label{fig:methodology}
\end{figure*}
In this STAR, we aim to adopt a wide lens on relevant research in the field of visualization literacy. We include research that deals with visualization literacy under other names (``graphicacy'', ``graph literacy'', etc.) as well as papers that tackle relevant concepts despite not explicitly mentioning visualization literacy in the title. Given the diversity of language used in these types of papers, we wanted to ground our paper collection methods in an examination of the literature. As such, we adopted an iterative approach to developing our survey methodology. Figure \ref{fig:methodology} provides an overview of the process. 

\subsection{Pilot Coding}
We first assembled a pilot corpus of 104 papers consisting of manually selected results from ``visualization literacy'' keyword searches on databases like IEEE Xplore, ACM Digital Library, and Web of Science, as well as known papers suggested by senior authors. The pilot corpus allowed us to iterate on the details of our methodology while also discussing examples of out-of-scope papers.

In order to define inclusion/exclusion criteria, one author conducted an informal open coding of the pilot corpus. Three initial categories of visualization literacy research emerged: evaluation (research that measures literacy), characterization (research that describes and models literacy), and intervention (research that improves literacy). In order to be considered relevant during this pilot stage, a paper must have made a contribution to visualization literacy in at least one of these categories. 

Two authors conducted a pilot round of coding using these criteria. Papers were tagged based on their contributions or marked for exclusion if no tag was applicable. Inter-coder reliability for paper inclusion was computed using Cohen's kappa, yielding a value of 0.633 (substantial agreement). A representative set of disagreement cases was discussed by all authors to clarify the definitions and limits of each category. We found that much of the disagreement stemmed from ambiguity around the characterization tag, which encompassed a wide range of contribution types. We later address this by expanding our initial framework into 5 categories, which also served as the final inclusion criteria (explained in section \ref{sect:inclusion}).

\begin{figure*}[htbp]
\centering
\includegraphics[width=\textwidth]{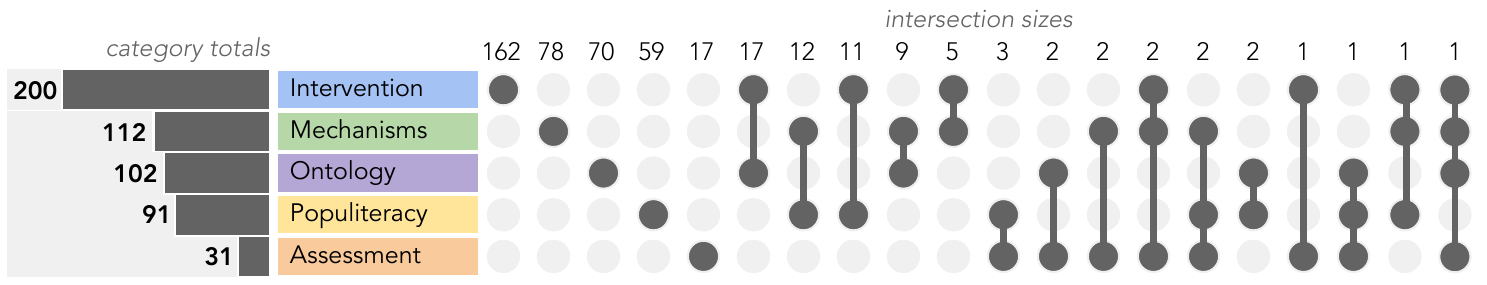}
\caption{The final survey corpus visualized in a modified UpSet plot. The bars on the left show the total number of papers in each category. Connected dots represent papers that fall under multiple categories, with the size of each intersection displayed as a number above the dots. \textbf{}}
\label{fig:upset}
\end{figure*}

\subsection{Keyword Search}
After the initial stage, we collected a set of potentially relevant papers using keyword search on five research databases: IEEE Xplore, ACM Digital Library, DBLP, Scopus, and Web of Science. The first three are computer science databases that cover IEEE Vis, ACM CHI, EuroVis, and other pertinent venues. Meanwhile, Web of Science and Scopus are general databases that index a broad range of research, allowing us to capture relevant work done outside of computer science (such as in psychology and education). 

We created our search term by manually extracting possible keywords from the titles and keyword lists of our pilot dataset. The resulting set of keywords included related concepts within visualization research, as well as other names used to refer to visualization literacy. We used the keywords in a boolean search term across all 5 databases:
\begin{verbatim}
(visualization AND (literacy OR education OR teach* OR learn* OR student OR novice)) 
OR 
(graphicacy OR "graph literacy" OR "graph comprehension" OR "graphical literacy" OR "graphical comprehension" OR "risk literacy")
\end{verbatim}

The structure and format of the search term was modified slightly for each database to account for variations in search functionality. Asterisks denote wildcard characters, while quotes force an exact match for the enclosed phrase. The quotes were necessary for terms like \textit{graph literacy}, where the word \textit{graph} could otherwise yield a large amount of results related to computer graphics or networks. This yielded 9,367 papers across all databases. After removing duplicate papers that were indexed by more than one database, 4,519 unique papers remained.

\subsection{Inclusion Coding}
\label{sect:inclusion}
Due to the subjective nature of assessing relevance to visualization literacy, we opted to systematically review each search result to determine inclusion. Based on discussions conducted during the pilot stage, we decided to expand our initial inclusion criteria from 3 categories to 5, reducing ambiguity between groups and achieving a more comprehensive scope. A paper was deemed relevant if it made a contribution to one or more of these research themes. The criteria are as follows:
\begin{itemize}
    \item \textbf{Assessment:} makes a contribution to the measurement of visualization literacy, for example by developing new test instruments.
    \item \textbf{Mechanisms:} models user processes related to literacy, such as learning strategies or barriers to understanding. 
    \item \textbf{Ontology:} addresses definitions, boundaries, and conceptualizations of visualization literacy (what visualization literacy \textit{is} and what it \textit{should be}).
    \item \textbf{Populiteracy:} describes the visualization literacy of a population of interest, such as a particular age group (e.g. children) or community (e.g. domain specialists). The name is a portmanteau of ``population'' and ``literacy''.
    \item \textbf{Intervention:} makes a contribution to the enhancement of visualization literacy skills by proposing and/or evaluating interventions.
\end{itemize}

The 4,519 search results were divided among 4 annotators. Each paper was tagged based on which categories it made contributions to; if no categories could be assigned, the paper was excluded. Throughout the process, the 4 annotators met regularly to ensure alignment on the definitions for each category. This process surfaced multiple examples of paper types to be excluded – for example, papers about ``spatial visualization'' (the cognitive ability to mentally manipulate shapes), or papers that were about the use of visualization in education (but not education about visualization).

Annotators also had the option to tag papers as ``contentious'' when uncertain about which tags to apply (or whether to apply any tags at all), allowing tags to be discussed by the group and then revisited. Once all papers were tagged, the remaining unresolved ``contentious'' papers were each reviewed by 2 new coders and included if at least 1 coder voted to do so. The result was a list of 406 papers that fit our inclusion criteria.

\subsection{Final Coding}
In the final coding stage, each of the included papers was reviewed by a new annotator. At this stage, each paper had been reviewed by at least 2 authors. In addition to assigning tags based on contribution, annotators also made note of the competencies and audiences addressed in each paper. In cases where the 2 authors assigned different tags, the paper in question was discussed until consensus was reached. An additional 53 papers were excluded during this process, leading to the set of 353 categorized papers. Finally, snowballing was conducted to include relevant forward and backward citations. The final survey corpus contains \textbf{374} relevant papers, each tagged with categories from our taxonomy, relevant competencies, and audiences. Figure \ref{fig:upset} shows how the papers are distributed across categories.

\section{Operationalizing Visualization Literacy through Competencies}
\label{sect:competencies}
In this section, we propose a \textbf{framework for operationalizing visualization literacy} using four competency themes that adapt based on the context of study. We arrived at the framework via a thorough analysis of the competencies addressed by papers in our survey corpus. We begin by motivating the importance of competencies, then explain each component of the framework. 

\begin{figure*}[htbp]
\centering
\includegraphics[width=\textwidth]{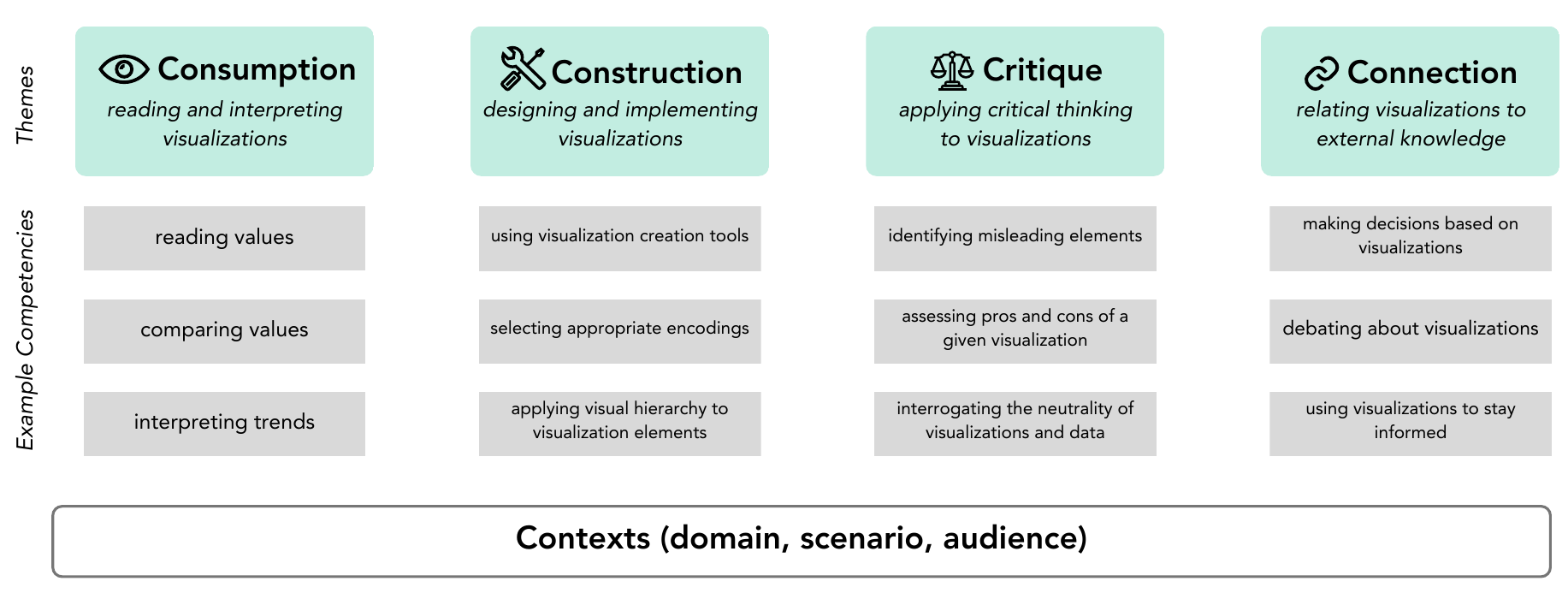}
\caption{A competency framework for visualization literacy. The four \textbf{competency themes} define the space of relevant visualization literacy competencies. Below each theme, example competencies illustrate some of the various levels of abstraction possible within each theme. \textbf{Contexts} cut across the themes, clarifying which aspects of which themes are relevant.}
\label{fig:competencies}
\end{figure*}

Various definitions of visualization literacy have been proposed in previous work, but most of them share a common thread: they focus on abilities, or ``competencies'', in the context of visualization. For example, consider the following three definitions: ``the ability to confidently use a given data visualization to translate questions specified in the data domain into visual queries in the visual domain, as well as interpreting visual patterns in the visual domain as properties in the data domain.'' \cite{BoyetalPrincipledWayAssessing2014}, ``the ability to confidently create and interpret visual representations of data'' \cite{AlperetalVisualizationLiteracyElementary2017}, and ``the ability to critically interpret and construct visualizations'' \cite{SolenScopingFutureVisualization2022}. These definitions encompass several different aspects of literacy, such as translation between visuals and data, interpretation, creation, and critical awareness. The diversity among them reflects differences in which competencies are emphasized and how they are framed. 

Visualization literacy research thus faces challenges in scoping and definition: it has been used by the community as a catch-all term for various related abilities, but an individual paper or study will usually only address a relevant subset of those abilities \cite{GeetalMoreComprehensiveUnderstanding2024}. As such, we argue that a useful definition of visualization literacy is one that clearly specifies which abilities are being addressed in the context where the definition is being used. We refer to the act of defining a relevant scope for visualization literacy based on context as \textbf{operationalizing visualization literacy}. Figure \ref{fig:competencies} illustrates our framework for creating these operationalizations. Our framework organizes the space of visualization literacy competencies into four themes: \textsc{consumption, construction, critique,} and \textsc{connection}. The relevant aspects of these themes vary depending on \textbf{context}, which involves factors such as the domain of study or the intended audience. Under each category, we list example competencies at various levels of abstraction. Rather than serve as an exhaustive list of competencies, however, the goal of the framework is to encourage more precise language when talking about visualization literacy. In the following paragraphs, we define each component of the framework and describe how to use the framework to operationalize visualization literacy for specific use cases.

\textsc{Consumption.} Competencies within this theme broadly deal with the ability to read and interpret visualizations. These include low-level tasks like reading or comparing values \cite{CarpenterShahModelPerceptualConceptual1998, LeeKimKwonVLATDevelopmentVisualization2017}, as well as higher-order skills like identifying patterns and trends \cite{RatwaniTraftonBoehm-DavisThinkingGraphicallyConnecting2008, QuadrietalYouSeeWhat2024}.

\textsc{Construction.} This theme deals with the knowledge and skills necessary to produce effective visualizations. Relevant competencies under this theme include designing visualizations \cite{FranconerietalScienceVisualData2021, AdelbergeretalIguanodonCodeBreakingGame2024, GeCuiKayAVECAssessmentVisual2025a} and implementing visualizations using code or other tools \cite{LoMingQuLearningVisTools2019, BattetalLearningTableauData2020, HedayatiKayChooseyourownD3Labs2023}. 

\textsc{Critique.} This theme covers how critical thinking abilities are applied to visualizations. Critical literacy in visualizations can range from identifying misleading elements of visualizations \cite{CambaCompanyByrdIdentifyingDeceptionCritical2022, GeCuiKayCALVICriticalThinking2023} to critically interrogating the processes involved in the production of visualizations \cite{GrayetalWaysSeeingData2016, ShreinerDataLiteracySocial2018}.

\textsc{Connection.} Competencies under this theme are about relating the data in a visualization to a broader context or external knowledge. These relations can happen internally (by forming opinions and mental models based on visualizations) \cite{GalesicGarcia-RetameroGraphLiteracyCrossCultural2011, LaietalMeasuringGraphComprehension2016} as well as externally (by debating about visualizations or using them to stay informed about relevant issues) \cite{PhilipOlivares-PasillasRochaBecomingRaciallyLiterate2016, DIgnazioCreativeDataLiteracy2022, BertlingHodgeRybaTheresPandemicGoing2023}.

\textbf{Contexts.} Cutting across the four themes, contexts refer to the situational factors that inform the relevance and scope of visualization literacy competencies. These include elements like the domain of use or the intended audience. For instance, visualization literacy research aimed at children may focus on lower-order \textsc{consumption} skills \cite{PanavasetalJuvenileGraphicalPerception2022} like the ability to correctly read values, whereas research addressing domain experts may be more interested in how their expertise is applied for decision-making (i.e. \textsc{connection}) \cite{JungjohannGebhardtScheerUnderstandingImprovingTeachers2022}.

\textbf{Themes, not categories.} A core feature of this competency framework is that the four themes are not disjoint categories, and competencies can thus be relevant to multiple themes. For example, health communication research has emphasized patients' ability to make informed lifestyle and medical decisions based on visualizations \cite{Garcia-RetameroCokelyCommunicatingHealthRisks2013, Garcia-RetameroCokelyDesigningVisualAids2017}. Doing so requires the \textsc{consumption} skills necessary to correctly read visualizations, as well as the \textsc{connection} skill of relating the presented information to one's own medical circumstances and well-being.

To illustrate how the framework can be used to operationalize visualization literacy in research, it may be helpful to reflect on some examples. We provide two hypothetical scenarios that demonstrate how operationalizations made with our framework can inform the scope and execution of a project.

Example 1: Consider a hypothetical study aimed at describing the visualization literacy of journalists through a series of interviews. For this study, visualization literacy would be defined as \textit{``the ability to critically interpret, effectively communicate with, and thoughtfully integrate data visualizations to enhance storytelling.''}

In this scenario, researchers can operationalize this definition through the framework's key competencies. The \textsc{critical} competencies come into play as journalists must evaluate visualizations for accuracy and potential bias, ensuring they don't mislead readers. \textsc{Construction} competencies are essential as journalists make design choices that effectively convey their story's data. Finally, \textsc{connection} competencies enable journalists to seamlessly integrate visualizations into their narratives, creating cohesive stories that leverage both text and visual elements. This framework-driven operationalization allows researchers to structure their interviews to explore how journalists develop and apply these specific visualization skills in their work.

Example 2: Consider a hypothetical research project called \textit{KinderVis} that aims to foster visualization literacy in kindergarteners. For this study, visualization literacy would be defined as \textit{``the foundational ability to understand, create, evaluate, and relate to basic data visualizations in ways that are developmentally appropriate for young children.''} This definition acknowledges both the specific needs of the target audience and the educational context of early childhood learning.

In this scenario, the researchers tailor their operationalization to the kindergarten classroom environment. \textsc{Consumption} competencies focus on building basic comprehension through simple bar and line graphs that represent familiar concepts like favorite colors or daily weather. \textsc{Construction} skills are developed through hands-on, age-appropriate activities like sketching and coloring their own simple charts. \textsc{Critical} thinking emerges as children learn to express their thoughts about visualizations, sharing what makes sense to them and what doesn't. \textsc{Connection} competencies are fostered by creating visualizations of classroom data - like favorite foods or playground activities - enabling children to see how data representations relate to their daily experiences and personal preferences. This framework-guided approach ensures that visualization literacy development aligns with both kindergarten learning capabilities and classroom teaching methods.

We encourage researchers to clearly define which aspects of visualization literacy they are addressing, using our framework as a guide. By reflecting on the application \textbf{context}, identifying relevant \textbf{competency themes}, and selecting appropriate \textbf{levels of abstraction}, researchers can tailor their definitions of visualization literacy to precisely articulate which competencies are most relevant to their work.

\section{Taxonomy of Visualization Literacy Research}
\label{sect:taxonomy}

Based on our iterative coding process, we identify 5 categories of visualization literacy research contributions: \onthl{ontology}, \asshl{assessment}, \mechhl{mechanisms}, \pophl{populiteracy}, and \inthl{intervention}. Figure \ref{fig:teaser} provides brief definitions for each category. We structure our discussion of the current literature around these categories, providing an overview of key concepts and trends for each. We also explore the interconnectedness of varied research contributions by discussing how advancements in each category inform and enable research in other categories. Each section concludes with a summary of key concepts, implications for other categories, and opportunities for future work.

\subsection{Ontology}
\label{sect:ontology}

Within our taxonomy of visualization literacy research, the ontology category includes work that approaches the concept at a meta level. This includes defining and scoping visualization literacy, highlighting relevant competencies, and describing its intersections with other literacies and contexts.

\subsubsection{Scoping Visualization Literacy}
A common thread among papers in this category is that they attempt to define the scope of visualization literacy by highlighting potentially neglected \textbf{competencies}. For example, multiple researchers have argued that the ability to identify misleading visualizations must be explicitly considered as a core element of visualization literacy \cite{CambaCompanyByrdIdentifyingDeceptionCritical2022, ChevalieretalObservationsReflectionsVisualization2018, GeCuiKayCALVICriticalThinking2023, SolenScopingFutureVisualization2022}. Beyond critical reading of visualizations, Gray et al. explore critical visualization literacy in terms of \textit{mediations} (e.g. between data and visualization, or between visualization and the reader), each of which merits critical inquiry and is imbued with meaning \cite{GrayetalWaysSeeingData2016}. Philip et al. particularly note how in the classroom setting, visualizations ``are embedded within structural and ideological relationships of power that make them replete with potential for productive (and unproductive) controversy'' \cite{PhilipOlivares-PasillasRochaBecomingRaciallyLiterate2016}.

Other ontological research in visualization literacy has sought to situate visualization literacy alongside other \textbf{individual traits}. He et al. introduce the concept of information receptivity, which is an individual's openness to receive data and interpretations of data. As a concept related to but distinct from visualization literacy, information receptivity demonstrates that ``individuals’ capacity for understanding and connecting with data reflects more than just the ability to read it'' \cite{HeetalEnthusiasticGroundedAvoidant2023}. Similarly, Mansoor and Harrison argue for the parallel study of visualization literacy alongside visualization biases by highlighting research establishing the links between cognitive ability, visualization experience, and susceptibility to bias \cite{MansoorHarrisonDataVisualizationLiteracy2018}.

Visualization literacy has also been situated in a variety of \textbf{application contexts}. In justifying the need for formal visualization education, Rushmeier et al. detail the various nontechnical professions that can benefit from visualization skills, such as history, archaeology, public policy, and literature \cite{RushmeieretalRevisitingNeedFormal2007}. Bobkowski and Etheridge conducted interviews with journalists and practitioners to elicit essential data skills for journalism students, finding support for teaching the ability to tell stories with data, as well as the ability to decide when visualizations are appropriate to use \cite{BobkowskiEtheridgeSpreadsheetsSoftwareStorytelling2023}. J{\"a}nicke describes how textual visualization skills can enable digital humanities researchers to not only gain insights into their data, but also to participate in interdisciplinary collaborations \cite{JanickeTeachingIntersectionVisualization2020}. More recently, an emerging strand of research has investigated how visualization literacy can be improved in audiences with varying ability levels and neurodivergence \cite{AlzahranietalUnlockingAutisticChildrens2024, SeoetalDesigningBornAccessibleCourses2024}. Other contexts where visualization literacy has been deemed important include biochemistry \cite{SchonbornAndersonImportanceVisualLiteracy2006}, social studies \cite{ShreinerDataliterateCitizenryHow2020}, and business \cite{KongExtendedAbstractWhat2020}.

\subsubsection{Complementary Literacies}
\label{sec:complementary}
Visualization literacy intersects with related literacies in various ways. The relationship between visualization literacy and other literacies is complementary: fluency in visualizations is a key component in many domains, and visualization literacy can also be informed and supported by disciplinary knowledge.

The most obvious example of visualization literacy supporting other literacies is data literacy. Pedersen and Cavaglia defined data literacy as ``a compound competence consisting  of some level of competence in statistics, data visualization and more generic competencies in problem-solving using different data'' \cite{PedersenCavigliaDataLiteracyCompound2019}, while Shreiner emphasized visualization's role in communicating discipline-specific information as a key component of data literacy \cite{ShreinerDataLiteracySocial2018}. Within data literacy, visualization literacy plays a communicative role (in sharing and understanding findings) as well as an analytic role (since insights during analysis can be supported/uncovered by visual methods) \cite{GrillenbergerRomeikeDevelopingTheoreticallyFounded2018}. 

In relation to scientific literacy, Gilbert notes that public engagement in science requires fluency in producing public displays of information \cite{GilbertRepresentationsModels2013}. Tang and Moje go so far as to include visualizations in their definition of all disciplinary literacies within science: ``the cultural practices that encompass specific ways of talking, writing, viewing, drawing, graphing, and acting, within a specialized discourse community'' \cite{TangMojeRelatingMultimodalRepresentations2010}. Other literacies which include visualization literacy as a component are information literacy \cite{WomackDataVisualizationInformation2014} and ecological literacy \cite{DobrinMoreyMediatingNatureRole2019}. 

Beyond acting as a necessary component of other literacies, visualization literacy plays an active role in shaping risk literacy and health literacy research. In risk literacy, defined as ``the ability to accurately interpret and make good decisions on the basis of information about risk'' \cite{Garcia-RetameroCokelyCommunicatingHealthRisks2013}, visual aids are used to teach patients about the benefits and drawbacks of certain medical procedures or to communicate flood risk in vulnerable communities \cite{RamasubramanianetalFloodRiskLiteracy2019}. Risk literacy research has found that visual aids can be used to overcome low numeracy in patients, as long as they also have moderate-to-high visualization literacy. However, overall low visualization literacy scores have also motivated a push for simplifying visual aid designs \cite{Garcia-RetameroCokelyDesigningVisualAids2017}. Other studies have argued that visualization literacy test scores should be prioritized over user preference when selecting between visualization types \cite{VanWeertetalPreferenceUnderstandingGraphs2021}. Lastly, Turchioe and Mangal include visualization literacy as a core component of patient-centered care, advising health practitioners to take into account visualization ability levels as well as cultural differences in the interpretations of graphics when communicating health information to patients \cite{ReadingTurchioeMangalHealthLiteracyNumeracy2024}.

The notion of complementary literacies is a two-way street, as concepts and skills from other literacies can be relevant to engaging with visualizations. For example, data literacy competencies such as data acquisition and cleaning can support the \textsc{construction} of visualizations, and critical evaluation of sources (a key aspect of information literacy) can also help in the process of \textsc{critiquing} the trustworthiness of a visualization. However, we note that the role of other literacies in visualization literacy has received limited attention. More explicit inquiry into which aspects of other literacies can support visualization literacy, both in research and in competency-building, is an open area of study. A clearer picture of useful concepts from other literacies may enable new modes of cross-disciplinary collaboration.

\subsubsection{Visualization Literacy as Meaning Making}
The myriad contexts of research and practice that intersect with visualization literacy speak to a broader point about the field: beyond merely a set of practical skills, visualization literacy is a gateway to making sense of and participating in the world around us. Intersections with other literacies demonstrate that visualization literacy can mediate access to information that people need to make crucial decisions about their lives or to stay informed about issues that directly affect them. Hedayati et al. referred to the social, cultural, and physical dimensions of visualization literacy as ``visualization practices'' \cite{HedayatiHuntKayPixelsPracticesReconceptualizing2024}, but otherwise, little attention has been given to this aspect of visualization literacy.

Visualization literacy's role in societal participation should influence how it is studied and taught. For example, Bertling et al. explore the potential of visualization as ``place-based education'' by arguing that visualization should be taught using data and contexts that directly connect to students' lived experiences. In doing so, students begin to view data more holistically and readily engage with it \cite{BertlingHodgeRybaTheresPandemicGoing2023}. Relatedly, D'Ignazio asserts that in order to teach data literacy to ``non-technical'' audiences, educators should embrace community-centered data and creative outputs like sketches and data sculptures \cite{DIgnazioCreativeDataLiteracy2022}. Multiple researchers have called for visualization literacy teaching to explicitly discuss the power structures and ethical issues in which visualizations are embedded, drawing attention to visualization's potential for harm if not critically examined \cite{ShreinerMartellMakingRaceRacism2024, PhilipOlivares-PasillasRochaBecomingRaciallyLiterate2016, KongExtendedAbstractWhat2020, McHenryUpdatingCodeTeaching2024}.

Despite these recognitions of visualization literacy's broader societal roles, current definitions fail to capture this importance. Chevalier et al. argue that visualization researchers are uniquely positioned to demonstrate the importance of visualization literacy and advocate for necessary improvements in visualization education \cite{ChevalieretalObservationsReflectionsVisualization2018}. We hope that this overview of visualization literacy as a process of meaning-making is a first step towards these goals.
\begin{tcolorbox}[title={Summary, Implications, and Open Areas}, ontbox]

Ontological research in visualization literacy explores meta-level aspects, including defining, scoping, and situating visualization literacy in broader contexts. Researchers emphasize overlooked aspects of visualization literacy, such as critical thinking about misleading visualizations or application contexts like journalism, digital humanities, and accessibility. Additionally, visualization literacy intersects with other literacies like data, scientific, and risk literacy, serving a complementary role in understanding and communicating complex information. 
\\

Concepts within this category have wide-ranging implications for other categories of visualization literacy research. In particular, competencies are the lifeblood of literacy research and can dictate the content of \asshl{assessment} questions, inform inquiry into the \mechhl{mechanisms} underlying various competencies, and enable more comprehensive \inthl{intervention} design. Recognition of underserved audiences in the context of visualization literacy could also motivate \pophl{populiteracy} studies. 
\\

Future research in this category should continue to interrogate the boundaries of visualization literacy and the roles it plays in various settings. Many studies emphasize practical skills but overlook visualization literacy's potential as a tool for meaning-making and societal participation. Holistic approaches to teaching visualization advocate for integrating lived experiences, ethical considerations, and power dynamics into education. Further exploration of how other literacies can support visualization literacy and its role in fostering critical societal engagement represents a key open area for future research.
\end{tcolorbox}

\subsection{Assessment}
\label{sect:assessment}
Research in the assessment category aims to reliably measure particular operationalizations of visualization literacy. The most common type of contribution in this category is a new instrument for assessing specific competencies. It is worth noting that in many of these papers, the operationalization of which competencies are being measured is implicit. As such, even if an assessment's stated goal is to measure ``visualization literacy'' as a whole, the competencies measured likely represent a specific operationalization of visualization literacy. Table \ref{table:asstable} summarizes papers that contribute assessments, the competencies tested in each, the assessment format, and the number of items. 

\begin{table*}[!h]\centering
\renewcommand\arraystretch{1.5} 
\caption{Summary of visualization literacy assessments}
\begin{tabular}{L{3cm} p{0.2cm} p{0.2cm} p{0.2cm} p{0.2cm} p{6cm} p{4cm} p{0.75cm}}
\hline
\textbf{Assessment} & \rotatebox{60}{\textsc{Consumption }} & \rotatebox{60}{\textsc{Construction }} & \rotatebox{60}{\textsc{Critique }} & \rotatebox{60}{\textsc{Connection }} & Specific Competencies Tested & Test Format & Items \\ \hline
\cite{BoyetalPrincipledWayAssessing2014} (Boy et al.) & {\large \textbullet} & & & & Retrieving values, making estimations, making comparisons & Multiple choice questions & 8-10 \\\hline
\cite{LeeKimKwonVLATDevelopmentVisualization2017} (VLAT) & {\large \textbullet} & & & & Retrieving values, making comparisons, finding correlations & Multiple choice questions & 53 \\\hline
\cite{CuietalAdaptiveAssessmentVisualization2023} (A-VLAT) & {\large \textbullet} & & & & Retrieving values, making comparisons, finding correlations & Multiple choice questions & 27 \\\hline
\cite{CuietalPromisesPitfallsUsing2025} (VILA-VLAT) & {\large \textbullet} & & & & Retrieving values, making comparisons, finding correlations & Multiple choice questions & 53 \\\hline
\cite{PandeyOttleyMiniVLATShortEffective2023} (Mini-VLAT) & {\large \textbullet} & & & & Retrieving values, making comparisons, finding correlations & Multiple choice questions & 12 \\\hline
\cite{FiratetalPLiteStudyParallel2022} (P-Lite) & {\large \textbullet} & & & & Retrieving values and identifying correlations in parallel coordinate plots & Multiple choice questions & 28 \\\hline
\cite{FiratDenisovaLarameeTreemapLiteracyClassroomBased2020} (Treemap Test) & {\large \textbullet} & & & & Reading and interpreting treemaps & Multiple choice questions & 27-30 \\\hline
\cite{FiratDenisovaLarameeTreemapLiteracyClassroomBased2020} (Treemap Rubric) & & {\large \textbullet} & & & Use of hierarchy, labels, node size mappings to create treemaps & Rubric applied on student-generated treemaps & 18 \\\hline
\cite{AdelbergeretalIguanodonCodeBreakingGame2024} (Iguanodon) & & {\large \textbullet} & & & Reduce overplotting, reduce ``chartjunk'', improve readability, use color encodings & Multiple choice questions & 12 \\\hline
\cite{GeCuiKayCALVICriticalThinking2023} (CALVI) & & & {\large \textbullet} & & Identifying misleading elements & Multiple choice questions & 30 \\\hline
\cite{CuietalAdaptiveAssessmentVisualization2023} (A-CALVI) & & & {\large \textbullet} & & Identifying misleading elements & Multiple choice questions & 15 \\\hline
\cite{GalesicGarcia-RetameroGraphLiteracyCrossCultural2011} (Objective Graph Literacy Scale) & {\large \textbullet} & & & {\large \textbullet} & Retrieving values, making comparisons, decision-making based on health visualizations & Numeric response, multiple choice & 10 \\\hline
\cite{Garcia-RetameroetalMeasuringGraphLiteracy2016} (Subjective Graph Literacy Scale) & {\large \textbullet} & & & {\large \textbullet} & Self-perception of abilities in retrieving values, making comparisons, decision-making based on health visualizations & Numeric scales & 5-10 \\\hline
\cite{OkanetalUsingShortGraph2019} (Short Graph Literacy Scale) & {\large \textbullet} & & & {\large \textbullet} & Retrieving and comparing values, decision-making based on health visualizations & Numerical response, multiple choice & 4 \\\hline
\cite{LaietalMeasuringGraphComprehension2016} (Lai et al.) & {\large \textbullet} & {\large \textbullet} & {\large \textbullet} & {\large \textbullet} & Reading features, interpreting relationships, sketching line graphs, comparing merits of visualizations, drawing conclusions about scientific phenomena & Multiple choice, short response, sketching & 14+ \\\hline
\end{tabular}
\label{table:asstable}
\renewcommand\arraystretch{1} 
\end{table*}

\subsubsection{Motivations for Testing}

The development of visualization literacy assessments is driven by various motivating factors. In a research context, assessments can be used to identify literacy gaps and develop more targeted interventions. Educators can also benefit from standardized ways of measuring student learning and evaluating the effectiveness of curricula. More broadly, insights into the comprehension abilities of target audiences can be useful for practitioners looking to effectively communicate information using visualizations.

The design and structure of assessments varies based on the \textbf{intended use case}. For example, the visualization literacy scale developed by Galesic and Garcia-Retamero is meant to assess visualization comprehension in relation to health risks and patient decision-making \cite{GalesicGarcia-RetameroGraphLiteracyCrossCultural2011}. In order for the test to be usable in clinical settings, they limited its length to 10 minutes and generated multiple-choice test items relevant to medical decision-making contexts. Meanwhile, Firat et al. developed an assessment intended to measure treemap literacy in a classroom setting \cite{FiratDenisovaLarameeTreemapLiteracyClassroomBased2020}. Their assessment takes the form of a rubric applied to student outputs from a treemap assignment; the rubric consists of ``yes'' or ``no'' questions about the correct use of treemap features like hierarchy and color mapping. These works are examples of how test design decisions are made to serve an (implicit or explicit) operational definition of visualization literacy relevant to the researchers and/or their context of use.

\subsubsection{Competencies Tested} 
A central question for assessment design and application is what competencies are being measured. Because visualization literacy encompasses many related but distinct competencies, it is useful to reflect on which skills are actually measured by each assessment. In Table \ref{table:asstable}, we summarize the competencies addressed by visualization literacy assessments in terms of the competency themes from Figure \ref{fig:competencies}.

The majority of existing visualization literacy assessments measure aspects of visualization \textsc{consumption}. Notable exceptions include the assessments developed for use in a clinical context, which all involve some level of \textsc{connection} competency in the context of health decision-making \cite{GalesicGarcia-RetameroGraphLiteracyCrossCultural2011, Garcia-RetameroetalMeasuringGraphLiteracy2016, OkanetalUsingShortGraph2019}. CALVI and its adaptive variant A-CALVI focus explicitly on \textsc{critique}, specifically the ability to identify misleading visualization elements \cite{GeCuiKayCALVICriticalThinking2023, CuietalAdaptiveAssessmentVisualization2023}. \textsc{Construction} has been measured both using multiple choice questions about visualization design (as in Iguanodon) or by applying a rubric to student output (as in the treemap assessment) \cite{AdelbergeretalIguanodonCodeBreakingGame2024, FiratDenisovaLarameeTreemapLiteracyClassroomBased2020}. One assessment, proposed by Lai et al., manages to cover aspects of all 4 themes. It was developed to measure visualization literacy in the context of K-12 science education, with a focus on integrating knowledge from visualizations into broader scientific concepts \cite{LaietalMeasuringGraphComprehension2016}.

\subsubsection{Test Construction}

A major factor in the development of assessments is the use of visualization tasks or processes to inform question creation. Lee et al. created the VLAT questions by compiling existing task taxonomies from information visualization, then mapping tasks to different visualization types. The matrix between tasks and visualizations formed a test blueprint from which items were constructed \cite{LeeKimKwonVLATDevelopmentVisualization2017}. Cui et al. generated their own task list using Bloom's taxonomy levels as a starting point \cite{CuietalPromisesPitfallsUsing2025}. Synthesizing multiple relevant taxonomies can also aid comprehensiveness, as was done by Ge et al., who merged two existing categorizations of misleading visualization elements to map out a design space for critical thinking questions in visualization \cite{GeCuiKayCALVICriticalThinking2023}. In the future, the development of new task taxonomies and process models may enable the assessment of various aspects of visualization literacy.

During the test construction process, more items are usually generated than are included in the final assessment. Pilot tests can be conducted using this larger set of items, after which the results are analyzed using psychometric methods to determine which ones are eventually included. A benefit of having a large number of test items is that they can comprise an item bank, or a set of items that test administrators can draw upon to customize the test's content or length \cite{BoyetalPrincipledWayAssessing2014, GeCuiKayCALVICriticalThinking2023, CuietalAdaptiveAssessmentVisualization2023}. Item banks provide additional flexibility in the application of assessments, allowing them to be tailored to specific use cases.

Because generating test items is time-consuming, automated item generation methods may prove useful when combined with human review. This was the approach taken by Cui et al. when developing the VILA pipeline, which uses large language models to generate visualization test items based on permutations of tasks, chart types, and data contexts \cite{CuietalPromisesPitfallsUsing2025}. A test modeled after VLAT tasks but composed of items from the VILA pipeline showed moderate to high convergent validity with the original test.

\subsubsection{Test Duration}

In many assessment use cases, it is ideal for tests to be administered as quickly as possible. The benefits of shorter assessments include preventing test fatigue, quickly making decisions based on the test result (i.e. in clinical settings), and the ability to administer the tests at scale. Assessment developers have adopted various strategies for reducing test duration while minimizing trade-offs in test validity or comprehensiveness. One approach has been to modify existing tests by systematically selecting a subset of items. Cui et al. draw on principles of computerized adaptive testing (CAT) to create shortened versions of two existing tests, VLAT and CALVI \cite{CuietalAdaptiveAssessmentVisualization2023}. By adaptively selecting the next question with the highest information based on item response theory, A-VLAT and A-CALVI each achieved high and acceptable reliability while using half the number of questions. In a similar vein, Pandey and Ottley shortened the VLAT using item-total correlation from a replication study of the VLAT in addition to validity ratings from visualization experts. As a result, their Mini-VLAT contains a 12-item subset of the original 53 items that achieves similar validity and reliability \cite{PandeyOttleyMiniVLATShortEffective2023}. 

Other strategies for test brevity involve the construction of the test items themselves, such as by having individual test items measure multiple skills at once \cite{AdelbergeretalIguanodonCodeBreakingGame2024}. Subjective self-rating of ability is also a possible solution when the use case allows for it, such as in clinical scenarios where the priority is to rapidly determine whether or not to use a visual aid with a patient \cite{Garcia-RetameroetalMeasuringGraphLiteracy2016}. While effective in reducing test duration, each of these approaches requires weighing of trade-offs in comprehensiveness of content or reduced rigor.

\subsubsection{Selecting the Appropriate Assessment} 
\label{subsec:selecting-assessments}
The previous subsections have demonstrated that the space of visualization literacy assessments varies in terms of competencies, tasks, chart types, test structure, and duration. As such, when looking to apply an existing assessment, care should be taken to ensure it suits the intended use case. 

Recent work has highlighted the need to ensure assessments are tailored to their context of use. For instance, {\"O}ney et al. interviewed domain experts who took the Mini-VLAT, finding that the 25-second time limit per question introduced stress factors which could obscure their actual ability level \cite{OneyetalTestingTestObservations2024}. The interviewees also felt that the Mini-VLAT questions focused too much on lower-order reading skills and would have benefited from questions connected to their domain of expertise. In a similar vein, Hedayati and Kay conducted interviews with university students before and after taking a university-level visualization course. They found that students exhibited increased ability to reason about visualizations from the designer's perspective. However, there was no significant effect on students' VLAT scores before and after the class, highlighting the disconnect in competencies taught in class (such as design empathy or deconstructing a given visualization into its component parts) versus what was being measured in the VLAT (basic \textsc{consumption} competencies) \cite{HedayatiKayWhatUniversityStudents2025}. These results dispel the notion of a ``one-size-fits-all'' test, emphasizing the importance of selecting an appropriate assessment for a given scenario.

To select the right test, it is crucial to consider both the competencies being measured and the test's duration. Table \ref{table:asstable} may serve as a useful starting point to determine whether the competencies measured in an assessment match the intended use case. For crowdsourced studies, a time limit or other shortening strategy may be a trade-off worth making, but for smaller sample sizes or more collaborative settings (such as with domain experts) it may be worth removing the time limit. Other considerations when selecting a test include whether the test is customizable or whether an item bank is needed for readministering the test.

\begin{tcolorbox}[title={Summary, Implications, and Open Areas}, assbox, breakable]
    Visualization literacy assessments span various use cases, such as health decision-making, classroom teaching, and user research. As a result, assessments can vary in terms of the competencies covered, test format, and duration, depending on the specific needs of the target audience. The majority of assessments, however, measure \textsc{consumption}-related competencies such as reading and interpreting. When administering assessments, it is important to consider whether the competencies measured (and other characteristics) are appropriate for the intended use case. 
    \\

    Assessments see widespread application in other areas of visualization literacy research. Existing assessments can be straightforwardly applied to \pophl{populiteracy} studies, while analysis of assessment results has been conducted to elicit \mechhl{barriers to literacy} \cite{NobreetalReadingPixelsInvestigating2024}. Assessments can also be used to evaluate the efficacy of visualization literacy \inthl{interventions} by measuring mastery of relevant concepts \cite{AdelbergeretalIguanodonCodeBreakingGame2024}.
    \\

    Future work in visualization literacy assessment may benefit from a greater diversity in test format. In particular, the lack of instruments for assessing \textsc{construction} competencies may be more appropriately addressed by rubric-based assessments rather than multiple choice. Scaling these rubric-based assessments is an open challenge which may benefit from automation. Further attention also needs to be given towards neglected competencies like deconstruction and design empathy, both of which fall under more than one competency theme. 
\end{tcolorbox}

\subsection{Mechanisms}
\label{sect:mechanisms}

The papers in our corpus categorized as Mechanisms includes \textbf{work that models user processes related to visualization literacy}, such as learning strategies or barriers to understanding. Understanding these underlying processes is crucial for developing effective interventions and teaching strategies \cite{CarpenterShahModelPerceptualConceptual1998}. 

From the analysis of research on mechanisms, three main topics emerge: (1) \textit{Cognitive models of visualization comprehension} (2) \textit{Individual differences}, and (3) \textit{Barriers to visualization literacy}. Research on cognitive models for visualization literacy investigates how humans process, interpret, and extract meaning from visual data representations through stages of perceptual encoding, pattern recognition, and semantic understanding. Research in individual differences investigates the role of cognitive and perceptual traits in visualization literacy. Lastly, barriers to visualization literacy includes research on the cognitive, educational, and design-related challenges that impede effective interpretation of data visualizations, including unfamiliarity with chart conventions and misalignment between visual encodings and perceptual capabilities.

\subsubsection{Cognitive Models of Visualization Comprehension}

Studies have revealed that visualization comprehension involves multiple interrelated cognitive processes operating at different levels of abstraction. One influential framework, proposed by Carpenter and Shah, emphasizes the critical role of pattern-to-interpretation integration cycles in visualization comprehension \cite{CarpenterShahModelPerceptualConceptual1998}. Their model outlines three key processing stages that occur cyclically: first, viewers identify and form chunks of visual information through pattern recognition. This is followed by two interpretative stages where integration occurs: viewers translate visual patterns into quantitative and qualitative meanings, including comparing spatial relationships and performing calculations, and then connect these interpreted patterns to the actual variables and concepts represented in the graph. Each cycle of integration builds upon previous cycles to construct a complete understanding of the visualization.

The effectiveness of these integration processes depends heavily on viewers' metacognitive strategies. Research has identified distinct levels of strategic sophistication in visualization comprehension, from basic to advanced. At the foundational level, Curcio's framework describes three progressive stages of engagement: reading literal values directly from the visualization, interpolating relationships between visible data points, and extrapolating beyond the displayed data to make predictions \cite{curcioComprehensionMathematicalRelationships1987}. The implementation of these strategies varies significantly with expertise. In their meta-analysis, Gegenfurtner et al. found that expert viewers employ more systematic search patterns and dedicate more attention to task-relevant visualization elements compared to novices \cite{GegenfurtnerLehtinenSaljoExpertiseDifferencesComprehension2011}. However, even experienced viewers can encounter difficulties. Trafton and Trickett identified common strategic errors in visualization comprehension, particularly when viewers persist with familiar but unsuitable approaches or fail to validate their initial interpretations \cite{TrickettTraftonComprehensiveModelGraph2006}.

Ratwani and colleagues built upon this foundation by distinguishing between two fundamental types of integration: visual integration, which involves recognizing patterns and forming visual groupings, and cognitive integration, where viewers engage in deeper reasoning about the visualization based on their specific goals or tasks \cite{RatwaniTraftonBoehm-DavisThinkingGraphicallyConnecting2008}. The cyclical nature of these processes is evident in viewers' eye movements, as they repeatedly scan between different graph elements such as the main display, axes, legends, and titles. 

Eye-tracking methodology has provided valuable insights into the temporal dynamics of visualization comprehension. Körner et al.'s research revealed that comprehension processes follow a sequential organization \cite{KorneretalEyeMovementsIndicate2014}, while Huestegge and Philipp demonstrated how spatial compatibility between visualization elements and legends can significantly reduce both comprehension time and the number of necessary gaze transitions \cite{HuesteggePhilippEffectsSpatialCompatibility2011}. Further research using eye-tracking methods has revealed that successful integration depends on both the type of visualization being viewed and individual user characteristics \cite{PeeblesAliExpertInterpretationBar2015}. For instance, the efficiency of connecting information between a visualization's legend and its data display can be significantly affected by design choices like direct labeling versus separate legends. This highlights how visualization comprehension emerges from an interaction between the viewer's capabilities and the visualization's design features.

\subsubsection{Individual Differences}

Research has identified several key user characteristics that influence how effectively people interpret visualizations, including visualization literacy skills, domain knowledge, familiarity with variables, and working memory capacity. Aguilar and Baek's research on student information-seeking preferences has added motivational factors to this list, highlighting how individual attitudes toward information gathering can impact visualization interpretation \cite{AguilarBaekMotivatedInformationSeeking2019}.
The relationship between attention and visualization comprehension has also received significant investigation. Fausett et al.'s research produced the intriguing finding that divided attention sometimes leads to improved performance in point estimation tasks \cite{FaussetRogersFiskUnderstandingRequiredResources2008}. In related work, Strobel et al. examined how seductive details – interesting but irrelevant elements – affect visualization comprehension, finding that while these elements increase processing time, they don't necessarily impair comprehension accuracy \cite{StrobelGrundLindnerSeductiveDetailsTheir2019}.

The role of spatial cognition has also emerged as a particularly crucial component in understanding visualization comprehension. Trickett and Trafton's research demonstrates that spatial processing becomes especially important in two key scenarios: when information isn't explicitly represented in the visualization, and when simple perceptual processes prove inadequate for extracting implicit information \cite{TrickettTraftonComprehensiveModelGraph2006}. This theoretical framework has found empirical support in studies like Stewart et al.'s work, which revealed that individuals with stronger spatial imaging abilities typically achieve higher levels of visualization comprehension \cite{StewartHunterBestRelationshipGraphComprehension2008}.
Eye-tracking methodology has provided valuable insights into the temporal dynamics of visualization comprehension. 

The format of visual presentations has also been shown to significantly influence comprehension processes. Research comparing different visualization methods, such as pie charts versus tree maps, has revealed performance benefits for rectangular representations in terms of both speed and accuracy \cite{HuesteggePotzschIntegrationProcessesFrequency2018}. Shah and Freedman's work further demonstrated that the choice between line graphs and bar graphs can lead viewers to different interpretations of the same data, particularly regarding interaction effects \cite{FreedmanShahModelKnowledgeBasedGraph2002}.



\subsubsection{Barriers to Visualization Literacy}

Understanding why people struggle with data visualizations has emerged as a critical research focus in recent years. While much work has focused on measuring and improving visualization literacy, researchers have increasingly turned their attention to understanding the fundamental barriers that prevent people from effectively comprehending visual representations of data \cite{FiratJoshiLarameeVisLitEVisualizationLiteracy2022}.

These barriers manifest across multiple dimensions, from basic comprehension challenges to deeper cognitive constraints. At the most fundamental level, many people struggle with the basic visual conventions that underpin data visualization. What might seem intuitive to experts—like understanding that position along an axis represents quantity or that connected lines imply continuity—actually requires explicit learning for novices \cite{ShahHoeffnerNoTitleFound2002}. This gap between expert assumptions and novice understanding can create significant obstacles to comprehension \cite{BrescianiEpplerPitfallsVisualRepresentations2015}.

The cognitive demands of processing visualizations present another significant challenge. When faced with complex visualizations containing multiple variables or relationships, viewers often encounter cognitive overload \cite{TrickettTraftonComprehensiveModelGraph2006}. Working memory limitations restrict how much information people can process simultaneously, leading them to focus selectively on certain aspects while potentially missing important insights \cite{FaussetRogersFiskUnderstandingRequiredResources2008}. This challenge becomes particularly acute when people encounter unfamiliar visualization types or need to understand complex data relationships \cite{NobreetalReadingPixelsInvestigating2024}.

 As one of the complementary literacies discussed in section \ref{sec:complementary}, statistical literacy plays a crucial role in visualization comprehension. Many people struggle with fundamental statistical concepts, which can severely impair their ability to interpret even relatively simple visualizations \cite{OkanetalBiasingDebiasingHealth2018}. This becomes especially problematic in contexts like healthcare, where misinterpreting probability or risk visualizations could have serious consequences. The intersection of statistical reasoning and visual interpretation creates a particularly challenging barrier for many viewers.

Research in museum settings has revealed that familiarity plays a crucial role in visualization comprehension. Visitors typically only succeed in understanding and identifying visualizations they've encountered before, while more complex representations, especially graph layouts, pose significant interpretative challenges \cite{BorneretalInvestigatingAspectsData2016}. This highlights how prior exposure and experience shape our ability to engage with different visualization types.

Socioeconomic factors also influence visualization literacy, with research revealing concerning disparities in visualization skills across different populations. These disparities are compounded by gaps in visualization literacy education within many educational systems, suggesting that broader systemic factors contribute to visualization literacy barriers \cite{GrammelToryStoreyHowInformationVisualization2010}. More recently, research has begun to organize these various challenges into clear categories: knowledge barriers (such as lack of statistical understanding), cognitive barriers (including working memory limitations), and sociocultural barriers (like lack of motivation or perceived relevance) \cite{NobreetalReadingPixelsInvestigating2024}.

When creating visualizations, users face additional challenges. Novices often struggle with selecting appropriate visualization types, understanding measurement semantics, and dealing with complex visual elements \cite{KwonFisherYiVisualAnalyticRoadblocks2011}. These creation-specific barriers highlight how visualization literacy involves not just passive comprehension, but also active engagement with data representation. For example, research on parallel-coordinates plots has identified specific challenges related to coordinate system understanding and visual confusion from intersecting lines \cite{PeeblesetalInfluenceGraphSchemas2013}.

This growing understanding of visualization literacy barriers has important implications for both design and education. Research suggests that adapted visualizations may particularly benefit those with lower visualization literacy, while feedback mechanisms can improve chart reading ability. However, addressing these barriers requires a comprehensive approach that considers not just the technical aspects of visualization but also the cognitive, educational, and social factors that influence how people engage with and understand visual representations of data.

\subsubsection{Competencies Studied}

Research on visualization literacy mechanisms reveals varying depths of understanding across the four core competency themes. While \textsc{consumption} competencies like basic visualization reading and interpretation have been extensively studied through cognitive and perceptual lenses, understanding of the other three competencies remains less developed.

Studies have revealed how viewers consume visualizations through pattern recognition and integration cycles \cite{CarpenterShahModelPerceptualConceptual1998}, with eye-tracking research demonstrating the sequential nature of comprehension \cite{KorneretalEyeMovementsIndicate2014} and how viewers navigate between visualization elements \cite{HuesteggePhilippEffectsSpatialCompatibility2011}.  Research demonstrates that even basic visualization tasks for \textsc{consumption} require sophisticated cognitive processes and can be impeded by various barriers, from fundamental difficulties with visual conventions \cite{ShahHoeffnerNoTitleFound2002} to cognitive overload \cite{TrickettTraftonComprehensiveModelGraph2006}.

Individual differences in factors like working memory capacity, spatial cognition, and domain knowledge significantly influence how people engage with visualizations across all competencies \cite{StewartHunterBestRelationshipGraphComprehension2008, AguilarBaekMotivatedInformationSeeking2019}. \textsc{Construction} competencies are primarily studied through research on visualization design choices, as demonstrated by studies comparing different visualization methods \cite{HuesteggePotzschIntegrationProcessesFrequency2018, FreedmanShahModelKnowledgeBasedGraph2002}.

In terms of \textsc{critique} competencies, viewers progress from basic comprehension to more systematic evaluation approaches \cite{curcioComprehensionMathematicalRelationships1987}, though this progression is heavily influenced by statistical literacy and domain knowledge \cite{OkanetalBiasingDebiasingHealth2018}. Studies of expert-novice differences have shown that experienced viewers employ more systematic search patterns and dedicate more attention to task-relevant elements \cite{GegenfurtnerLehtinenSaljoExpertiseDifferencesComprehension2011}. 

Research on \textsc{connection} competencies is exemplified by Ratwani's distinction between visual and cognitive integration, showing how viewers move from pattern recognition to deeper reasoning based on their goals and knowledge \cite{RatwaniTraftonBoehm-DavisThinkingGraphicallyConnecting2008}. 

The effectiveness of these processes across all competencies is mediated by both individual characteristics and contextual factors, including sociocultural barriers that affect how different populations engage with visualizations \cite{NobreetalReadingPixelsInvestigating2024}. However, while we have deep insights into consumption processes, our understanding of the mechanisms supporting creation, critique, and connection competencies remains limited, suggesting important directions for future research in visualization literacy.

\begin{tcolorbox}[title={Mechanisms: Summary, Implications, and Open Areas}, mechbox, breakable]

Research on visualization literacy mechanisms has revealed the complex cognitive processes underlying how people understand, create, and engage with data visualizations. Studies have identified multiple interrelated processes, from basic pattern recognition to sophisticated integration cycles that connect visual patterns with meaning.  Together, these studies paint a picture of visualization comprehension as a sophisticated cognitive process that depends heavily on individual capabilities, prior knowledge, and contextual factors.
\\

Understanding these mechanisms has important implications for visualization literacy \asshl{assessment} and \inthl{intervention} design. For example, knowledge of cognitive load barriers has led to recommendations for progressive complexity in visualization education \cite{WojtonetalBeginBeginningConstructionist2018}. Similarly, understanding of individual differences has informed the development of adaptive assessment tools and personalized learning approaches \cite{StoferCheComparingExpertsNovices2014}.
\\

While substantial research exists on basic visualization comprehension mechanisms, several critical areas remain underexplored. We need deeper investigation into the cognitive mechanisms underlying competencies beyond \textsc{consumption}, particularly in \textsc{construction}, \textsc{critique}, and \textsc{connection} with broader contexts. 
\end{tcolorbox}

\subsection{Populiteracy}
\label{sect:populiteracy}

Many visualization literacy studies made primary contributions of measuring and describing the visualization literacy of a population of interest, for example a particular age group or community. Hence the name ``populiteracy'', which is a portmanteau of ``population'' and ``literacy''.
Contributions include improving visualization literacy education through a better understanding of students' learning processes, improving data accessibility for vulnerable populations, enabling better decision-making in domain-specific contexts of visualization use, and characterizing the visualization literacy ability of ``the public'', which many research works refer to as a general audience for visualizations. 
We construct a taxonomy within these populiteracy papers of similar populations, such as ``Students'', encompassing for example K-12 as well as college-focused studies (see Table \ref{table:poptable}). 

\begin{table*}[t]
\label{table:poptable}
\centering
\caption{Populiteracy subcategories and included groups}
\begin{tabular}{L{4cm} @{\hskip 20pt} p{3.5cm} @{\hskip 20pt} p{7cm}}  
\hline
\textbf{Population} & \textbf{Grouping} & \textbf{Members} \\ \hline
\multirow{2}{*}{\textbf{Students}} & \vspace{0.25cm} Education & \begin{itemize}[left=0pt,label=--] \item Primary school students (elementary-, middle-, and high school) \item K-12 students (cuts across categories of primary schooling) \item Teachers \item Undergraduate students \item Graduate students \end{itemize} \\ \cline{2-3}
 & \vspace{0.25cm} Relevant Age Groups & \begin{itemize}[left=0pt,label=--] \item Children \item Adolescents \end{itemize} \\ \hline
\multirow{2}{*}{\textbf{Inclusive Design Populations}} & \vspace{0.25cm} Accessibility & \begin{itemize}[left=0pt,label=--] \item Visually impaired individuals \item People with dyslexia \end{itemize} \\ \cline{2-3}
 & \vspace{0.05cm} Relevant Age Group & \begin{itemize}[left=0pt,label=--] \item Older adults \end{itemize} \\ \hline
\multirow{2}{*}{\textbf{Domain Specialists}} & \vspace{0.25cm} Decision-Making & \begin{itemize}[left=0pt,label=--] \item Policymakers \item Heads of household \end{itemize} \\ \cline{2-3}
 & \vspace{0.25cm} Healthcare & \begin{itemize}[left=0pt,label=--] \item Nurses \item Physicians \end{itemize} \\ \hline
\multirow{2}{*}{\textbf{General Public}} & \vspace{0.25cm} Socioeconomic Status & \begin{itemize}[left=0pt,label=--] \item Unschooled individuals \item Individuals with lower socioeconomic standing \end{itemize} \\ \cline{2-3}
 & \vspace{0.25cm} General Population Samples & \begin{itemize}[left=0pt,label=--] \item Museum goers \item ``The public'' as an audience \item Geographical population sample \end{itemize} \\ \hline
\end{tabular}
\renewcommand\arraystretch{1} 
\end{table*}

\subsubsection{Students}

Studies targeting student populations motivated their measurements of visualization literacy by focusing on challenges in the education space. These included adjusting the school curriculum to teach visualization literacy skills, or identifying gaps in teacher understanding and support for teaching visualization literacy skills throughout the K-12 education program \cite{CernyPotancokInformationLiteracyInternational2023}. 
In particular, researchers have pointed to the need for graph literacy-specific training for teachers \cite{JungjohannGebhardtScheerUnderstandingImprovingTeachers2022}. 
Other studies have found that their target population possesses a ``medium'' level of literacy, in high school students for instance\cite{TiroRulianaAswiLiteracyDescriptionProbability2021}, and use these visualization literacy assessment results to call for better approaches to teaching visualization literacy alongside specific material, such as probability and statistics \cite{OzkanEsraArikanMehmetOzkanStudyVisualizationSkills2018}. 
Studies in education have also taken a broader view, for example examining ``when and how'' students acquire visualization literacy during their schooling \cite{MalteseHarshSvetinaDataVisualizationLiteracy2015}, 
and connecting visualization literacy to skills beyond doing well in school, such as how visualization literacy skills taught in school can impact daily adult life beyond the classroom \cite{RidingBoardmanRelationshipSexLearning1983}. 


Teachers themselves are also the target of some prior work on visualization literacy, with some studies measuring teacher beliefs and attitudes as well as making suggestions for how to approach visualization literacy in the classroom.
One study identified possible pitfalls in teaching visualization literacy, demonstrating that teachers may overlook data points critical to correct interpretation and do not always show a ``consistent strategy'' in teaching \cite{JungjohannGebhardtScheerUnderstandingImprovingTeachers2022}. 
Other work points out that teachers of undergraduate courses may overestimate the visualization literacy ability of their students \cite{MalteseHarshSvetinaDataVisualizationLiteracy2015}, and thus need to specifically teach students to read visualizations to be able to develop those skills.

Correlating student visualization literacy with other skills was another theme that emerged from the analysis. 
Cox et al. found that students had equal performance in visualization tasks that were non-subject specific relative to those that were subject specific (e.g. programming versus non-programming tasks) \cite{CoxetalCognitiveProcessingPerspective2004}. 
Ozkan et al. examined visualizations for abstract mathematical concepts, finding that some populations of students actually prefer not to use visualizations, instead performing better with rule-based reasoning instead \cite{OzkanEsraArikanMehmetOzkanStudyVisualizationSkills2018}.
Another study reports that no significant differences were found in visualization literacy ability between students with less STEM experience coursework than those with more\cite{MalteseHarshSvetinaDataVisualizationLiteracy2015}.
One overarching observation from these different studies is that they often use different definitions or measurement scales for visualization literacy, a potentially confounding factor worthy of further investigation.


Characterizing these studies in terms of visualization literacy \textbf{competencies} (Figure \ref{fig:competencies}), most of the studies included tasks relating to \textsc{consumption} \cite{CoxetalCognitiveProcessingPerspective2004} \cite{JungjohannGebhardtScheerUnderstandingImprovingTeachers2022} \cite{MalteseHarshSvetinaDataVisualizationLiteracy2015} \cite{CernyPotancokInformationLiteracyInternational2023} \cite{MalteseHarshSvetinaDataVisualizationLiteracy2015}, and to a lesser degree those related to \textsc{connection} \cite{CoxetalCognitiveProcessingPerspective2004} \cite{JungjohannGebhardtScheerUnderstandingImprovingTeachers2022}, although a few evaluated \textsc{construction} tasks \cite{OzkanEsraArikanMehmetOzkanStudyVisualizationSkills2018} \cite{MalteseHarshSvetinaDataVisualizationLiteracy2015}.







Prior work in this category uncovers several areas of need for future work.
Mokhtar et al., for example, recommend designing and teaching visualization literacy abilities to students in specific disciplines such as engineering \cite{AliMokhtarVisualizationSkillsUniversiti2014}. 
Improvements to recruitment and sampling strategies were also identified as helping strengthen the reliability and generalizability of findings.
For instance, Maltese et al. make the explicit choice of recruiting participants with varied levels of experience in the STEM fields \cite{MalteseHarshSvetinaDataVisualizationLiteracy2015}. 
Our analysis also reveals competencies that were underrepresented, such as the \textsc{Critique} category. As critical thinking is clearly an important skill for student populations, future work-- possibly leveraging recent advances in critical visualization literacy assessments (e.g. \cite{GeCuiKayCALVICriticalThinking2023}), could possibly fill this gap. 

\subsubsection{Inclusive Design Populations}

Inclusive design ensures users with a wide variety of characteristics, including age and accessibility-related characteristics, are represented in design processes and outcomes \cite{kendrickInclusiveDesign2022}. 
Throughout these studies, the importance of visualization literacy is often represented through the idea that visualizations are increasingly encountered on a daily basis, and therefore making them broadly accessible is an important dimension. 

Identified studies focus on the intersection of inclusivity and health or health risk visualizations.
For example, Shah et al. examine the intersection of medically-defined ``frailty''--adverse outcomes such as high morbidity, e.g. for older adults with advanced age-- and the influences of health and graph literacy \cite{ShahetalAssociationHealthLiteracy2019}. 
They found that higher visualization literacy was associated with a lower risk of frailty in older adults, suggesting that literacy levels may play a role in empowering people to managing their health outcomes.
Similarly, Feng et al. characterizes the data visualization literacy skills of young adults with Down Syndrome, towards the goal of improving their access to and ability to use health-related data visualizations \cite{WoodFengLazarHealthDataVisualization2024}.
They also point out that the (in)accessibility of health information could increase the adverse effects of inequity that these already more vulnerable populations face. 

Other studies measure visualization literacy in the context of specific ability characteristics, e.g. low vision, blindness, or dyslexia, with the goal of identifying or evaluating potential interventions.
Kim et al., for example, evaluate visualization literacy in people with dyslexia, finding increased response times for common chart-reading and comprehension tasks \cite{KimetalInvestigatingGraphComprehension2014}.
Similarly, Acarturk et al. develop a chart-reading assistant, and evaluate visualization literacy in a population of blind or impaired users in order to examine the potential benefit of their tool \cite{AcarturkAlacamHabelDevelopingVerbalAssistance2014}.
Common to these studies is the goal of evaluating visualization literacy in order to assess the potential negative impacts of impairments. 

\textbf{Competencies} evaluated in these studies fall under \textsc{Consumption} and \textsc{Connection} categories.
\textsc{Construction} remains an important open area, as inclusive design would plausibly include interfaces for the creation of visualizations as well as interfaces for reading them. Similarly, an examination of \textsc{Critique} may also reveal unique challenges in the ecosystem of techniques and interfaces for visualization users with specific impairments and needs.

Given the smaller number of papers in the space of visualization literacy and inclusive design, it is not surprising that existing work generally calls for more attention to the area. Shah et al., for example, call for longitudinal studies with populations that could potentially benefit from targeted research focus (e.g. veteran populations) \cite{ShahetalAssociationHealthLiteracy2019}.
Similarly, Feng et al., who worked with young adults with Down Syndrome, identify a need for additional studies to examine the generalizability of their results.
The proposed competency framework is one way to identify paths forward. For example, since the studies tended to focus on \textsc{consumption} and \textsc{connection}, further research might cover \textsc{construction} and \textsc{critique}, which may present unique benefits, such as creating or critiquing charts for more effective advocacy.



\subsubsection{Domain Specialists}

Studies also target visualization literacy in specific domains, examining how well people can use visualizations in specific vocations (e.g. nurses and physicians) or in areas of responsibility (policymakers or heads of households). 

Herrmann et al. investigated whether people could understand the links between their everyday decisions and the resulting data \cite{HerrmannBrumbyOreszczynWattsYourUsage2018}. 
Specifically, they measure how well people made decisions with visualizations with a given goal (such as lowering energy usage).
Barone et al. examine the link between visualization literacy as it pertains to financial information and resulting financial behavior \cite{BaroneBussoliFattobeneExploringFinancialGraph2024}, with specific recommendations for policymakers.
Herrmann et al. suggest that energy data visualizations are challenging to act on in the context of daily use \cite{HerrmannBrumbyOreszczynWattsYourUsage2018}, such that it may not be enough to focus on increasing the visualization literacy of the visualization audience, but that changes in the visual format would be required as well. 
One takeaway across these studies is the influence of routines of engagement with data and visualization, which may shape participants' process for coming to a given conclusion based on the visualizations. 

Health and medicine emerged as another discipline where visualization literacy measures were used to assess specific populations.
For example, Lopez et al. examine the visualization literacy of nurses, with the given goal of informing how dashboards and interfaces are designed \cite{LopezetalNursesNumeracyGraphical2016}. Their findings suggest that current design practices do not sufficiently integrate notions of visualization literacy, raising potential risks for software that is critical in healthcare settings.
Similarly, Rungvivatjarusa et al. report a study that evaluates physician visualization literacy in order to evaluate a specific tool of interest \cite{RungvivatjarusetalTrainingPediatricPhysicians2024}, finding that a tool designed with data visualization as a focal point led to increased reported confidence in using visualizations in important clinical decision-making tasks.




Across these studies, one salient theme was a need for increased attention and development of visualization literacy-related measures, tools, and evaluations. Such efforts were seen as potential means for advancing domain-specific interests, for instance better financial decision-making, and better and more efficient usage of health information in medical settings.

\subsubsection{General Public}

Studies focusing on the general public described two groups of populations: the first encompassing social characteristics such as individuals' educational standing (e.g. “unschooled participants” \cite{McKenziePadillaConstructionValidationTest1986}), and socioeconomic status \cite{DurandetalGraphLiteracyMatters2020}, while the second group spanned public data visualization users (e.g. museums), novices versus experts, and medical patients.
Many studies referred to the general public when motivating their respective contributions.
Interestingly, the definitions of what ``public'' means varied.
Examples of varying definitions include participants recruited from “public cafes and public libraries”, email, and social media \cite{HeetalEnthusiasticGroundedAvoidant2023}.
In contrast, another work defines public as sampling from local university staff, construction site employees, and local farmer's market visitors\cite{PeckAyusoEl-EtrDataPersonalAttitudes2019}.

Key findings on visualization literacy in the general public include that 1) it varies, and 2) that visualization techniques may also need to vary to align with the interests and abilities of populations.
For example, Peck et al. recommend that personal connection to data may override other design factors when it comes to graph perception \cite{PeckAyusoEl-EtrDataPersonalAttitudes2019}.
Shaffer et al., examining varying visualization designs in a general population, found that individuals with high graph literacy more accurately completed tasks with a smoothed graph than participants with lower graph literacy, highlighting that chart type itself may intersect with visualization literacy on performance measures \cite{ShafferetalPatientJudgmentsHypertension2022}. 
Findings from these studies also appear to align, with Peck et al. also reporting that rural area participants rated charts with simple visual encodings (such as bar and line graphs) as more useful to them \cite{PeckAyusoEl-EtrDataPersonalAttitudes2019}.


Studies examining visualization literacy for the general public also seemed to align on calls for future work with societal challenges.
Peck et al. call for research to counter misinformation by forming personal connections to data \cite{PeckAyusoEl-EtrDataPersonalAttitudes2019}. 
Similarly Shaffer et al. suggest conducting similar studies with physicians, to identify how design recommendations may change based on expertise.

\begin{tcolorbox}[title={Summary, Implications, and Open Areas}, popbox, breakable]

Populiteracy studies focus on characterizing the visualization literacy of a given group of individuals, whether in the education space, inclusive-design related populations, domain specialists, or the broader group of the general public (which is often used in this area of research). 
\\

One important category of populiteracy papers are the ones that use a visualization literacy \asshl{assessment} to characterize the ability levels in various competencies of a given population, ranging from the commonly-studied group consisting of students to more underrepresented individuals under the accessibility grouping. In doing so, they contribute factors to consider for improved \inthl{intervention} approaches to teaching visualization literacy. They also suggest possible \mechhl{mechanisms} through which individuals of a given group reason.
\\

Papers in this category have identified open areas for future work, such as verifying the generalizability of findings from one population to another \cite{WoodFengLazarHealthDataVisualization2024} and additional efforts to characterize the visualization literacy of students within engineering and other technical fields \cite{AliMokhtarVisualizationSkillsUniversiti2014}. In addition, future populiteracy work should examine competencies which were not covered by existing studies, especially \textsc{critique} and \textsc{construction}.

\end{tcolorbox}

\subsection{Intervention}
\label{sect:intervention}
The papers in the intervention category of our taxonomy address efforts to improve visualization literacy. These papers vary in approach, with many papers discussing pedagogical practices or educational approaches for teaching visualization in the classroom, either as an experience report (e.g. \cite{MaTeachingVisualization2005,WolfeTeachingStudentsFocus2015,LanReflectionsTeachingData2024,StrantzDataBrickolageTeaching2022}) or with a formal evaluation (e.g. \cite{WangVisVisualToolkitTeaching2022,MacedoetalBuildingInformationVisualization2022}). Others describe the design or implementation of non-classroom interventions designed to improve visualization literacy, including tools (e.g. \cite{StoiberetalDAnoNoLearningEnvironment2023}) and workshops (e.g. \cite{KejstovaetalConstructPlayEngaging2023}). In addition to papers focusing on specific interventions, a number of papers focused on surveying existing interventions (e.g. by surveying visualizations in textbooks \cite{PostigoLopez-ManjonGraphicacyBiologyTextbooks2015,ShreinerMartellMakingRaceRacism2024,Medley-RathetalFiguresChartsTables2024}) or reflecting on what aspects of visualization should be taught \cite{NolanPerrettTeachingLearningData2016,ChevalieretalObservationsReflectionsVisualization2018,CammMcCrayRoehmMoreJustCharts2023}, or surveying a variety of visualization classes \cite{KerrenInformationVisualizationCourses2013}. 

While some of the papers in our survey explicitly aim to improve visualization literacy, other papers implicitly improve visualization literacy within the context of teaching a visualization course or using visualizations in other teaching---these papers may not be framed specifically as being about visualization literacy. We include them here for a fuller picture of how visualizations are taught, whether explicitly as part of visualization literacy interventions or not. Additionally, while this survey focuses on research papers, many visualization literacy interventions may not be discussed by the research community. This may include books about data visualization aimed at a general audience \cite{KnaflicStorytellingDataData2015,SmithHowChartsWork2022}, online courses \cite{OnlineCourseAnalyzing,NJBCMakingBetter}, and articles about visualization design \cite{8TipsMake2020}.

The papers in the other categories of our taxonomy tended to focus more on \textsc{consumption} competencies such as the ability to read and interpret visualizations, whereas the majority of intervention papers focused on improving skills related to \textsc{construction} or \textsc{critique} of visualizations. Interventions related to improving \textsc{consumption} competencies tended to be about learning specific visualization types (e.g. bar charts \cite{AlperetalVisualizationLiteracyElementary2017}, network visualizations \cite{ZhouetalUsingNetworkVisualizations2024}, and parallel coordinates plots \cite{FiratetalPLiteStudyParallel2022}), and often also included other competencies.


\subsubsection{Intervention approaches}

In designing interventions to be more effective or engaging, a number of approaches were used. One common approach was to use \textbf{game-based tools or interventions}, which could increase engagement and motivation to learn \cite{PandelievetalSeriousGameTeaching2022}. These game-based interventions include Diagram Safari, an educational game to foster data literacy in children \cite{GableretalDiagramSafariVisualization2019}, Roboviz, a game-based final project for a data visualization class \cite{AdarLee-RobbinsRobovizGameCenteredProject2022}, Iguanodon, a game application in which students construct visualizations using design choices \cite{AdelbergeretalIguanodonCodeBreakingGame2024}, and a game for teaching data literacy to undergraduate students \cite{PandelievetalSeriousGameTeaching2022}.

Many interventions also focus on \textbf{the context of the data} as a way of engaging people. Some interventions engage participants by providing them with a fictional motivating scenario, such as that they ``have been hired as junior editors at the OhNo Gazette'' \cite{ChinBlairSchwartzGotGameChoiceBased2016} or that they are recruiting robots, and aim to ``recruit the most positively productive robots'' \cite{AdarLee-RobbinsRobovizGameCenteredProject2022}. Others use real-world data to give students a sense of authenticity in the tasks they do and to ``heighten the meaningfulness of the results'' \cite{NagelVisuallyAnalysingUrban2020}, such as a dataset about a locally familiar animal (mountain gazelles) \cite{BarYardenOhDeerPracticing2023} or a dataset about ``statistical indicators of people's living conditions'' \cite{StenlidenBodenNissenStudentsProducersInteractive2019}. Given that people gravitate towards visualizations that connect with their personal experiences, many interventions also use personally relevant data to ``stimulate curiosity and interest'' \cite{NagelVisuallyAnalysingUrban2020}, such as visualizing data collected together as a class \cite{StrantzDataBrickolageTeaching2022} or choosing locally meaningful data \cite{ZhouetalUsingNetworkVisualizations2024,LanReflectionsTeachingData2024}. Interventions such as Peppler et al.'s museum exhibition even explicitly give users tasks such as finding themselves in the data and comparing themselves to others \cite{PepplerKeuneHanCultivatingDataVisualization2021}.

Researchers and educators often design interventions based on \textbf{existing learning theory and pedagogical approaches} that enable learning through exploration and inquiry. This includes Kejstová et al.'s \cite{KejstovaetalConstructPlayEngaging2023} workshop, which uses a constructivist playful approach to increase students' understanding of visual encodings. Peppler et al. \cite{PepplerKeuneHanCultivatingDataVisualization2021} use Papert's constructionist theory of learning \cite{PapertMindstormsChildrenComputers1980} to design a museum exhibit in which visitors can ``engage with and make sense of contextualized data''. Other approaches include using Indigenous knowledge-making practices \cite{StrantzDataBrickolageTeaching2022}, active learning \cite{StoiberetalDAnoNoLearningEnvironment2023}, or inquiry-based learning \cite{NagarajRajaDhiliphanRajkumarDevelopmentInquirybasedActive2024} to help students learn through exploration.



\subsubsection{Interventions for \textsc{construction} competencies}
The majority of papers in the intervention category focused on constructing visualizations, including different aspects of the visualization creation pipeline: data skills, visualization design, and visualization implementation.

\textbf{Working with data. }
As discussed in \autoref{sec:complementary}, data literacy and visualization literacy are strongly connected, and many visualization literacy interventions therefore explicitly emphasize developing skills related to working with data. This can include identifying appropriate datasets \cite{LoMingQuLearningVisTools2019}, dealing with the messiness of authentic scientific data \cite{EllweinetalUsingRichContext2014}, and understanding characteristics of data such as data types and structures \cite{ByrdDwengerActivityWorksheetsTeaching2021}. Others focus on learning statistical and data analytic methods \cite{ByrdDwengerActivityWorksheetsTeaching2021}, or on alleviating students' ``fear of data'' \cite{LanReflectionsTeachingData2024}. Stoiber et al., for example, created a tool to guide data journalists through the data analytics process with a step-by-step approach \cite{StoiberetalDAnoNoLearningEnvironment2023}. 
Skills related to understanding data and statistics are often emphasized as a prerequisite to constructing visualizations from the data.

\textbf{Design thinking. }
Another important aspect of visualization design interventions is an emphasis on making choices in the visualization design process to effectively tell a story or communicate a message. One important aspect of visualization design and storytelling is understanding tradeoffs of different choices made during the design process, such as  selecting appropriate data or choosing visual encodings. Indeed, students in visualization classes do seem to get better at ``design empathy'', or the awareness of the choices made by the designer in the creation process \cite{HedayatiKayWhatUniversityStudents2025}. Wolfe \cite{WolfeTeachingStudentsFocus2015}, for example, describes a classroom activity in which students are given a dataset about the Olympics and are asked to create a data visualization that presents an assigned country ``in the best possible light while still representing the data ethically'', which helps students think about the choices that can be made to present a particular message. Similarly, Stenliden et al. \cite{StenlidenBodenNissenStudentsProducersInteractive2019} had students work in pairs to produce a visualized story using a real world dataset, while Ahmad \& Ma \cite{AhmadMaImpactsStudentLLM2024} had students work on a data storytelling exercise using data from the Titanic, giving students a goal of creating visualizations that are customized to a particular message or story.

Other interventions explicitly related to design choices include workshops for design ideation where people create visualizations using analog tools such as Lego, paper, and colored pencils \cite{StrantzDataBrickolageTeaching2022,KejstovaetalConstructPlayEngaging2023}, often with an emphasis on the design process rather than the finished product \cite{HeAdarVizItCardsCardBasedToolkit2017}. By encouraging a variety of visualizations of the same data, interventions such as He \& Adar's VizIt cards \cite{HeAdarVizItCardsCardBasedToolkit2017} create an explicit opportunity for learners to think about the variety of choices that could be made in visualization design. Given the value of prototyping and flexibility in visualization design, other tools such as Spence et al.'s creative data visualization tool \cite{SpenceetalIncreasingDataKnowledgeArtistic2021} lower technical barriers by allowing users to create visualizations through artistic representations. Similarly, Bishop et al. designed a free-form visualization tool called Construct-A-Vis \cite{BishopetalConstructAVisExploringFreeForm2020} to allow children to create visualizations without being restricted to a particular chart type.


\textbf{Implementation. }
While some interventions abstracted away the technical process of creating visualizations, others specifically focused on teaching people how to use particular technical tools to implement visualizations. Lo et al. \cite{LoMingQuLearningVisTools2019} describe a set of tutorials for using tools such as Tableau, Python and JavaScript to create visualizations. Hedayati \& Kay \cite{HedayatiKayChooseyourownD3Labs2023} describe an assignment for teaching students to adapt online D3 code, while Batt et al. \cite{BattetalLearningTableauData2020} describe an assignment for teaching fundamental Tableau concepts to undergraduate students.

\subsubsection{Interventions for other competencies}
While the other categories of our taxonomy had a large number of papers related to \textsc{consumption} competencies such as reading and interpreting visualizations, interventions solely focusing on these skills were less common, except for interventions which introduce specific new visualization types. Additionally, interventions often emphasized other competencies, such as the ability to identify deceptive visualizations and the ability to deconstruct a visualization into its constituent parts.


\textbf{Reading specific visualizations. }
Interventions about reading and interpreting visualizations tended to be within introductions of specific chart types. For example, Alper et al. \cite{AlperetalVisualizationLiteracyElementary2017} surveyed elementary school textbooks and found three categories of exercises: reading exercises, completion exercises, and construction exercises. Because elementary schoolers are usually being explicitly taught new visualization types, these reading exercises make sense to include in conjunction with exercises about other competencies. Similarly, in introducing network visualizations to fifth and sixth graders, Zhou et al. \cite{ZhouetalUsingNetworkVisualizations2024} teach students to extract information from a network visualization in addition to activities related to tasks like data collection. 

Other interventions introduce less frequently used visualization types, such as interventions teaching people to read and use treemaps \cite{FiratetalConstructivismbasedApproachTreemap2023,FiratDenisovaLarameeTreemapLiteracyClassroomBased2020,FuchsetalTreEducationVisualEducation2024}. While these interventions use a variety of approaches, including having students construct and manipulate treemaps, one key learning outcome is for them to be able to read and interpret them accurately. Interventions for specific visualization types also include tools to teach parallel coordinates plots \cite{FiratetalPLiteStudyParallel2022}, and teaching students to use biochemistry visualizations such as ball-and-stick representations \cite{SchonbornAndersonImportanceVisualLiteracy2006}.

\textbf{Identifying deceptive visualizations. }
In terms of critical thinking and deceptive visualization, there are some interventions that are about teaching people the ability to identify misleading or poorly constructed visualizations. Hopkins et al. \cite{HopkinsCorrellSatyanarayanVisuaLintSketchySitu2020}, for example, introduce a tool called VisuaLint to highlight chart construction errors, and find that users more reliably identified these errors after being shown examples using VisuaLint. Visualization classes also often explicitly talk about design flaws and reasoning flaws to help students identify them \cite{LanReflectionsTeachingData2024}. Camba et al. compared different interventions for teaching students about deceptive practices: in-class discussion, a computer-based tool, and a hands-on activity in which students created deceptive visualizations. While the goal of this intervention was to help students identify deceptive visualization practices when viewing visualizations, the construction based intervention was most effective in doing so.

\textbf{Deconstructing visualizations. }
Another competency that some interventions address is the ability to deconstruct a visualization into its component parts \cite{HedayatiKayWhatUniversityStudents2025}. 
Amalbili et al.'s \textit{Guess-Vis?} game \cite{AmabiliGuptaRaidouTaxonomyDrivenModelDesigning2021} focuses on deconstructing visualizations by having a player guess the identity of a visualization by asking questions about its characteristics, which requires players to identify characteristics of a visualization. Adelberger et al.'s Iguanodon game \cite{AdelbergeretalIguanodonCodeBreakingGame2024} similarly emphasizes  deconstruction by explicitly highlighting particular actions that can be taken (e.g. changing the grid lines or changing the mark type). Alper et al. \cite{AlperetalVisualizationLiteracyElementary2017} introduce a tool for teaching elementary schoolers about visualizations using a concreteness fading approach in which they transition from a more concrete visualization (pictographs) to a more abstract visualization (bar charts), and by doing so, they help students understand the structure and components behind the chart type.


\begin{tcolorbox}[title={Summary, Implications, and Open Areas}, intbox, breakable]
    Intervention papers include papers about teaching visualization in a classroom context, either as part of a data visualization class or incorporated in another classroom setting. This category also includes papers about one-off interventions such as workshops or museum exhibits, as well as the development of tools and games intended to increase visualization literacy. The majority of interventions focus on \textsc{construction} related competencies, with a smaller number focused on reading and interpreting specific visualization types, identifying deceptive visualizations, and deconstructing visualizations. \\

    A number of intervention papers focus on  \onthl{ontological} themes such as which competencies are part of visualization literacy and which we should design interventions for. Intervention papers are also highly related to \mechhl{mechanism} papers, as we need to understand the processes underlying visualization literacy in order to design effective interventions. Similarly, as we better understand the visualization literacy of \pophl{particular groups}, we can design better interventions to meet the needs of different populations. \\   

    Future research about visualization literacy interventions could look into existing interventions outside a classroom context and their effectiveness, including corporate workshops, guidelines on visualization design best practices, and articles aimed at a general audience. Additionally, more work could focus on how to generalize existing interventions to new contexts.

\end{tcolorbox}

\section{Discussion and Conclusion}
\label{sect:conclusion}

Our survey reveals where work in visualization literacy 
has concentrated and highlights gaps in the literature. In this section, we discuss high-level takeaways about visualization literacy more broadly.

\textbf{Gaps in competency coverage.} By operationalizing visualization literacy in terms of competencies throughout this paper, we revealed differential coverage of the four competencies across categories of research.
For example, existing \asshl{assessments} and \pophl{populiteracy} studies primarily measure \textsc{consumption} competencies, while literacy \inthl{interventions} have focused on teaching the \textsc{construction} of visualizations. Addressing underexplored competencies, both in terms of the four themes and in terms of level of abstraction within those themes, creates opportunity for future work. There is also a need to investigate higher-level competencies that draw on multiple themes, such as design empathy and deconstruction (\autoref{subsec:selecting-assessments}), which involve abilities across \textsc{consumption}, \textsc{construction}, and \textsc{critique}. Further investigation into how multiple competency themes compound together could lead to exciting new research directions.

\textbf{Future work should precisely operationalize visualization literacy.} Research surveyed in this report has covered a broad space of operationalizations of visualization literacy. Too often, however, these operationalizations are implicit, causing ambiguity about what competencies are being addressed in a given paper, or how that work relates to broader notions of visualization literacy. Researchers should precisely express which competencies they aim to address when studying visualization literacy in order to situate their work in the broader literature and enable more consistent comparisons across studies. We hope that our framework for operationalizing visualization literacy (\autoref{sect:competencies}), including contexts of use and the four competency themes, will help researchers more precisely define their research scope. At the same time, we expect that expanding and refining this framework---lending the framework itself more precision---will be a fruitful area of future work.

\textbf{Under-studied populations.} While there is an increasing body of work dedicated to describing the visualization literacy of \pophl{particular groups}, there are several populations that merit further study. We echo Solen's observation that the vast majority of visualization literacy research is conducted in a Western context \cite{SolenScopingFutureVisualization2022}. Furthermore, little is understood about how to approach visualization literacy for people with disabilities such as visual impairments. As fluency in visualization becomes increasingly necessary to participate in society, visualization literacy researchers are uniquely positioned to improve equitable access to information, and part of that task is ensuring that no group of people is left behind.

Through a comprehensive survey of visualization literacy---both inside and outside the visualization domain---we have provided an overview of research categories and explored how they connect to visualization competencies. Visualization literacy is an exciting and constantly evolving field that has the potential to reshape how we think about and interact with data. We hope that our STAR serves as a useful reference for key concepts in visualization literacy and inspires solutions to open areas of research. 

\bibliographystyle{eg-alpha-doi} 
\bibliography{STARZoteroManual}       

\newcommand{\etalchar}[1]{$^{#1}$}
\begin{thebibliography}{\uppercase{VWAVDJ21}}

\bibitem[8Ti20]{8TipsMake2020}
8 tips to make better charts.
\newblock \url{https://qz.com/work/1869894/8-tips-to-improve-the-your-chart-and-graph-design}, June 2020.

\bibitem[AAH14]{AcarturkAlacamHabelDevelopingVerbalAssistance2014}
\textsc{Acart{\"u}rk C., Ala{\c c}am {\"O}., Habel C.}:
\newblock Developing a {{Verbal Assistance System}} for {{Line Graph Comprehension}}.
\newblock In \emph{Design, {{User Experience}}, and {{Usability}}. {{User Experience Design}} for {{Diverse Interaction Platforms}} and {{Environments}}}, Hutchison D., Kanade T., Kittler J., Kleinberg J.~M., Kobsa A., Mattern F., Mitchell J.~C., Naor M., Nierstrasz O., Pandu~Rangan C., Steffen B., Terzopoulos D., Tygar D., Weikum G., Marcus A., (Eds.), vol.~8518. Springer International Publishing, Cham, 2014, pp.~373--382.
\newblock \url{http://link.springer.com/10.1007/978-3-319-07626-3_34}.
\newblock \href {https://doi.org/10.1007/978-3-319-07626-3_34} {\path{doi:10.1007/978-3-319-07626-3_34}}.

\bibitem[AB19]{AguilarBaekMotivatedInformationSeeking2019}
\textsc{Aguilar S.~J., Baek C.}:
\newblock Motivated {{Information Seeking}} and {{Graph Comprehension Among College Students}}.
\newblock In \emph{Proceedings of the 9th {{International Conference}} on {{Learning Analytics}} \& {{Knowledge}}} (Tempe AZ USA, Mar. 2019), ACM, pp.~280--289.
\newblock \url{https://dl.acm.org/doi/10.1145/3303772.3303805}.
\newblock \href {https://doi.org/10.1145/3303772.3303805} {\path{doi:10.1145/3303772.3303805}}.

\bibitem[AC01]{AdcockCollierMeasurementValidityShared2001}
\textsc{Adcock R., Collier D.}:
\newblock Measurement {{Validity}}: {{A Shared Standard}} for {{Qualitative}} and {{Quantitative Research}}.
\newblock \emph{American Political Science Review 95}, 3 (Sept. 2001), 529--546.
\newblock \url{https://www.cambridge.org/core/journals/american-political-science-review/article/measurement-validity-a-shared-standard-for-qualitative-and-quantitative-research/91C7A9800DB26A76EBBABC5889A50C8B}.
\newblock \href {https://doi.org/10.1017/S0003055401003100} {\path{doi:10.1017/S0003055401003100}}.

\bibitem[AGR21]{AmabiliGuptaRaidouTaxonomyDrivenModelDesigning2021}
\textsc{Amabili L., Gupta K., Raidou R.~G.}:
\newblock A {{Taxonomy-Driven Model}} for {{Designing Educational Games}} in {{Visualization}}.
\newblock \emph{IEEE Computer Graphics and Applications 41}, 6 (Nov. 2021), 71--79.
\newblock \url{https://ieeexplore.ieee.org/document/9556564/}.
\newblock \href {https://doi.org/10.1109/MCG.2021.3115446} {\path{doi:10.1109/MCG.2021.3115446}}.

\bibitem[AL22]{AdarLee-RobbinsRobovizGameCenteredProject2022}
\textsc{Adar E., {Lee-Robbins} E.}:
\newblock Roboviz: {{A Game-Centered Project}} for {{Information Visualization Education}}.
\newblock \url{https://arxiv.org/abs/2208.04403}, 2022.
\newblock \href {https://doi.org/10.48550/ARXIV.2208.04403} {\path{doi:10.48550/ARXIV.2208.04403}}.

\bibitem[ALE{\etalchar{*}}24]{AdelbergeretalIguanodonCodeBreakingGame2024}
\textsc{Adelberger P., Lesota O., Eckelt K., Schedl M., Streit M.}:
\newblock Iguanodon: {{A Code-Breaking Game}} for {{Improving Visualization Construction Literacy}}.
\newblock \emph{IEEE Transactions on Visualization and Computer Graphics} (2024), 1--14.
\newblock \url{https://ieeexplore.ieee.org/document/10697307/}.
\newblock \href {https://doi.org/10.1109/TVCG.2024.3468948} {\path{doi:10.1109/TVCG.2024.3468948}}.

\bibitem[AM14]{AliMokhtarVisualizationSkillsUniversiti2014}
\textsc{Ali D.~F., Mokhtar M.}:
\newblock Visualization skills among {{Universiti Teknologi Malaysia}} student.
\newblock In \emph{2014 {{International Symposium}} on {{Technology Management}} and {{Emerging Technologies}}} (Bandung, Indonesia, May 2014), IEEE, pp.~139--142.
\newblock \url{http://ieeexplore.ieee.org/document/6936494/}.
\newblock \href {https://doi.org/10.1109/ISTMET.2014.6936494} {\path{doi:10.1109/ISTMET.2014.6936494}}.

\bibitem[AM24]{AhmadMaImpactsStudentLLM2024}
\textsc{Ahmad M., Ma K.-L.}:
\newblock Impacts of {{Student LLM Usage}} on {{Creativity}} in {{Data Visualization Education}}.
\newblock \emph{EuroVis 2024 - Education Papers} (2024).
\newblock \url{https://diglib.eg.org/handle/10.2312/eved20241055}.
\newblock \href {https://doi.org/10.2312/EVED.20241055} {\path{doi:10.2312/EVED.20241055}}.

\bibitem[ARC{\etalchar{*}}17]{AlperetalVisualizationLiteracyElementary2017}
\textsc{Alper B., Riche N.~H., Chevalier F., Boy J., Sezgin M.}:
\newblock Visualization {{Literacy}} at {{Elementary School}}.
\newblock In \emph{Proceedings of the 2017 {{CHI Conference}} on {{Human Factors}} in {{Computing Systems}}} (Denver Colorado USA, May 2017), ACM, pp.~5485--5497.
\newblock \url{https://dl.acm.org/doi/10.1145/3025453.3025877}.
\newblock \href {https://doi.org/10.1145/3025453.3025877} {\path{doi:10.1145/3025453.3025877}}.

\bibitem[ATJW24]{AlzahranietalUnlockingAutisticChildrens2024}
\textsc{Alzahrani M.~S., Tag B., Johnson B., Wybrow M.}:
\newblock Unlocking {{Autistic Children}}'s {{Potential}}: {{The Crux}} with {{Data Visualisations}} and {{IoT}}.
\newblock In \emph{Companion of the 2024 on {{ACM International Joint Conference}} on {{Pervasive}} and {{Ubiquitous Computing}}} (Melbourne VIC Australia, Oct. 2024), ACM, pp.~701--705.
\newblock \url{https://dl.acm.org/doi/10.1145/3675094.3678477}.
\newblock \href {https://doi.org/10.1145/3675094.3678477} {\path{doi:10.1145/3675094.3678477}}.

\bibitem[Bal76]{BalchinGraphicacy1976}
\textsc{Balchin W. G.~V.}:
\newblock Graphicacy.
\newblock \emph{The American Cartographer 3}, 1 (Jan. 1976), 33--38.
\newblock \url{https://www.tandfonline.com/doi/full/10.1559/152304076784080221}.
\newblock \href {https://doi.org/10.1559/152304076784080221} {\path{doi:10.1559/152304076784080221}}.

\bibitem[BBF24]{BaroneBussoliFattobeneExploringFinancialGraph2024}
\textsc{Barone M., Bussoli C., Fattobene L.}:
\newblock Exploring financial graph literacy: Determinants and influence on financial behavior.
\newblock \emph{Qualitative Research in Financial Markets} (Oct. 2024).
\newblock \url{https://www.emerald.com/insight/content/doi/10.1108/QRFM-12-2023-0304/full/html}.
\newblock \href {https://doi.org/10.1108/QRFM-12-2023-0304} {\path{doi:10.1108/QRFM-12-2023-0304}}.

\bibitem[BBG19]{BornerBueckleGindaDataVisualizationLiteracy2019}
\textsc{B{\"o}rner K., Bueckle A., Ginda M.}:
\newblock Data visualization literacy: {{Definitions}}, conceptual frameworks, exercises, and assessments.
\newblock \emph{Proceedings of the National Academy of Sciences 116}, 6 (Feb. 2019), 1857--1864.
\newblock \url{https://pnas.org/doi/full/10.1073/pnas.1807180116}.
\newblock \href {https://doi.org/10.1073/pnas.1807180116} {\path{doi:10.1073/pnas.1807180116}}.

\bibitem[BCHH23]{BachetalVisualizationEmpowermentHow2023}
\textsc{Bach B., Carpendale S., Hinrichs U., Huron S.}:
\newblock \emph{Visualization {{Empowerment}}: {{How}} to {{Teach}} and {{Learn Data Visualization}} ({{Dagstuhl Seminar}} 22261)}.
\newblock Tech. Rep.~6, Schloss Dagstuhl -- Leibniz-Zentrum f{\"u}r Informatik, 2023.
\newblock \url{https://drops.dagstuhl.de/entities/document/10.4230/DagRep.12.6.83}.
\newblock \href {https://doi.org/10.4230/DAGREP.12.6.83} {\path{doi:10.4230/DAGREP.12.6.83}}.

\bibitem[BD21]{ByrdDwengerActivityWorksheetsTeaching2021}
\textsc{Byrd V., Dwenger N.}:
\newblock Activity {{Worksheets}} for {{Teaching}} and {{Learning Data Visualization}}.
\newblock \emph{IEEE Computer Graphics and Applications 41}, 6 (Nov. 2021), 25--36.
\newblock \url{https://ieeexplore.ieee.org/document/9547790/}.
\newblock \href {https://doi.org/10.1109/MCG.2021.3115396} {\path{doi:10.1109/MCG.2021.3115396}}.

\bibitem[BE15]{BrescianiEpplerPitfallsVisualRepresentations2015}
\textsc{Bresciani S., Eppler M.~J.}:
\newblock The {{Pitfalls}} of {{Visual Representations}}: {{A Review}} and {{Classification}} of {{Common Errors Made While Designing}} and {{Interpreting Visualizations}}.
\newblock \emph{Sage Open 5}, 4 (Oct. 2015), 2158244015611451.
\newblock \url{https://journals.sagepub.com/doi/10.1177/2158244015611451}.
\newblock \href {https://doi.org/10.1177/2158244015611451} {\path{doi:10.1177/2158244015611451}}.

\bibitem[BE23]{BobkowskiEtheridgeSpreadsheetsSoftwareStorytelling2023}
\textsc{Bobkowski P.~S., Etheridge C.~E.}:
\newblock Spreadsheets, {{Software}}, {{Storytelling}}, {{Visualization}}, {{Lifelong Learning}}: {{Essential Data Skills}} for {{Journalism}} and {{Strategic Communication Students}}.
\newblock \emph{Science Communication 45}, 1 (Feb. 2023), 95--116.
\newblock \url{https://journals.sagepub.com/doi/10.1177/10755470221147887}.
\newblock \href {https://doi.org/10.1177/10755470221147887} {\path{doi:10.1177/10755470221147887}}.

\bibitem[BGHT20]{BattetalLearningTableauData2020}
\textsc{Batt S., Grealis T., Harmon O., Tomolonis P.}:
\newblock Learning {{Tableau}}: {{A}} data visualization tool.
\newblock \emph{The Journal of Economic Education 51}, 3-4 (Sept. 2020), 317--328.
\newblock \url{https://www.tandfonline.com/doi/full/10.1080/00220485.2020.1804503}.
\newblock \href {https://doi.org/10.1080/00220485.2020.1804503} {\path{doi:10.1080/00220485.2020.1804503}}.

\bibitem[BHH{\etalchar{*}}21]{BachetalSpecialIssueVisualization2021}
\textsc{Bach B., Huron S., Hinrichs U., Roberts J.~C., Carpendale S.}:
\newblock Special {{Issue}} on {{Visualization Teaching}} and {{Literacy}}.
\newblock \emph{IEEE Computer Graphics and Applications 41}, 6 (Nov. 2021), 13--14.
\newblock \url{https://ieeexplore.ieee.org/document/9646527/}.
\newblock \href {https://doi.org/10.1109/MCG.2021.3117412} {\path{doi:10.1109/MCG.2021.3117412}}.

\bibitem[BHR23]{BertlingHodgeRybaTheresPandemicGoing2023}
\textsc{Bertling J.~G., Hodge L., Ryba E.}:
\newblock ``{{There}}'s a {{Pandemic Going On}}!'': {{Data Visualization}} as {{Critical Place-Based Education}} in {{Challenging Times}}.
\newblock \emph{Art Education 76}, 3 (May 2023), 24--31.
\newblock \url{https://www.tandfonline.com/doi/full/10.1080/00043125.2023.2167156}.
\newblock \href {https://doi.org/10.1080/00043125.2023.2167156} {\path{doi:10.1080/00043125.2023.2167156}}.

\bibitem[BMBH16]{BorneretalInvestigatingAspectsData2016}
\textsc{B{\"o}rner K., Maltese A., Balliet R.~N., Heimlich J.}:
\newblock Investigating aspects of data visualization literacy using 20 information visualizations and 273 science museum visitors.
\newblock \emph{Information Visualization 15}, 3 (July 2016), 198--213.
\newblock \url{https://journals.sagepub.com/doi/10.1177/1473871615594652}.
\newblock \href {https://doi.org/10.1177/1473871615594652} {\path{doi:10.1177/1473871615594652}}.

\bibitem[BRBF14]{BoyetalPrincipledWayAssessing2014}
\textsc{Boy J., Rensink R.~A., Bertini E., Fekete J.-D.}:
\newblock A {{Principled Way}} of {{Assessing Visualization Literacy}}.
\newblock \emph{IEEE Transactions on Visualization and Computer Graphics 20}, 12 (Dec. 2014), 1963--1972.
\newblock \url{http://ieeexplore.ieee.org/document/6875906/}.
\newblock \href {https://doi.org/10.1109/TVCG.2014.2346984} {\path{doi:10.1109/TVCG.2014.2346984}}.

\bibitem[BY23]{BarYardenOhDeerPracticing2023}
\textsc{Bar C., Yarden A.}:
\newblock Oh {{Deer}} {\dots} {{Practicing Scientific Inquiry}} and {{Data Literacy}} through an {{Authentic Gazelle Data Set}}.
\newblock \emph{The American Biology Teacher 85}, 5 (May 2023), 245--251.
\newblock \url{https://online.ucpress.edu/abt/article/85/5/245/196213/Oh-Deer-Practicing-Scientific-Inquiry-and-Data}.
\newblock \href {https://doi.org/10.1525/abt.2023.85.5.245} {\path{doi:10.1525/abt.2023.85.5.245}}.

\bibitem[BZP{\etalchar{*}}20]{BishopetalConstructAVisExploringFreeForm2020}
\textsc{Bishop F., Zagermann J., Pfeil U., Sanderson G., Reiterer H., Hinrichs U.}:
\newblock Construct-{{A-Vis}}: {{Exploring}} the {{Free-Form Visualization Processes}} of {{Children}}.
\newblock \emph{IEEE Transactions on Visualization and Computer Graphics 26}, 01 (Jan. 2020), 451--460.
\newblock \url{https://www.computer.org/csdl/journal/tg/2020/01/08807271/1cG66gYAFtS}.
\newblock \href {https://doi.org/10.1109/TVCG.2019.2934804} {\path{doi:10.1109/TVCG.2019.2934804}}.

\bibitem[CBS16]{ChinBlairSchwartzGotGameChoiceBased2016}
\textsc{Chin D.~B., Blair K.~P., Schwartz D.~L.}:
\newblock Got {{Game}}? {{A Choice-Based Learning Assessment}} of {{Data Literacy}} and {{Visualization Skills}}.
\newblock \emph{Technology, Knowledge and Learning 21}, 2 (July 2016), 195--210.
\newblock \url{http://link.springer.com/10.1007/s10758-016-9279-7}.
\newblock \href {https://doi.org/10.1007/s10758-016-9279-7} {\path{doi:10.1007/s10758-016-9279-7}}.

\bibitem[CCB22]{CambaCompanyByrdIdentifyingDeceptionCritical2022}
\textsc{Camba J.~D., Company P., Byrd V.}:
\newblock Identifying {{Deception}} as a {{Critical Component}} of {{Visualization Literacy}}.
\newblock \emph{IEEE Computer Graphics and Applications 42}, 1 (Jan. 2022), 116--122.
\newblock \url{https://ieeexplore.ieee.org/document/9693369/}.
\newblock \href {https://doi.org/10.1109/MCG.2021.3132004} {\path{doi:10.1109/MCG.2021.3132004}}.

\bibitem[CGD{\etalchar{*}}23]{CuietalAdaptiveAssessmentVisualization2023}
\textsc{Cui Y., Ge L.~W., Ding Y., Yang F., Harrison L., Kay M.}:
\newblock Adaptive {{Assessment}} of {{Visualization Literacy}}.
\newblock \url{https://arxiv.org/abs/2308.14147}, 2023.
\newblock \href {https://doi.org/10.48550/ARXIV.2308.14147} {\path{doi:10.48550/ARXIV.2308.14147}}.

\bibitem[CGD{\etalchar{*}}25]{CuietalPromisesPitfallsUsing2025}
\textsc{Cui Y., Ge L.~W., Ding Y., Harrison L., Yang F., Kay M.}:
\newblock Promises and {{Pitfalls}}: {{Using Large Language Models}} to {{Generate Visualization Items}}.
\newblock \emph{IEEE Transactions on Visualization and Computer Graphics 31}, 1 (Jan. 2025), 1094--1104.
\newblock \url{https://ieeexplore.ieee.org/document/10670418/}.
\newblock \href {https://doi.org/10.1109/TVCG.2024.3456309} {\path{doi:10.1109/TVCG.2024.3456309}}.

\bibitem[CHRA{\etalchar{*}}18]{ChevalieretalObservationsReflectionsVisualization2018}
\textsc{Chevalier F., Henry~Riche N., Alper B., Plaisant C., Boy J., Elmqvist N.}:
\newblock Observations and {{Reflections}} on {{Visualization Literacy}} in {{Elementary School}}.
\newblock \emph{IEEE Computer Graphics and Applications 38}, 3 (May 2018), 21--29.
\newblock \url{https://ieeexplore.ieee.org/document/8370203/}.
\newblock \href {https://doi.org/10.1109/MCG.2018.032421650} {\path{doi:10.1109/MCG.2018.032421650}}.

\bibitem[cKFY11]{KwonFisherYiVisualAnalyticRoadblocks2011}
\textsc{chul Kwon B., Fisher B., Yi J.~S.}:
\newblock Visual analytic roadblocks for novice investigators.
\newblock In \emph{2011 {{IEEE Conference}} on {{Visual Analytics Science}} and {{Technology}} ({{VAST}})} (Oct. 2011), pp.~3--11.
\newblock \url{https://ieeexplore.ieee.org/document/6102435}.
\newblock \href {https://doi.org/10.1109/VAST.2011.6102435} {\path{doi:10.1109/VAST.2011.6102435}}.

\bibitem[CMR23]{CammMcCrayRoehmMoreJustCharts2023}
\textsc{Camm J.~D., McCray G.~E., Roehm M.~L.}:
\newblock More than just charts and graphs: {{What}} to teach in a data visualization course.
\newblock \emph{Decision Sciences Journal of Innovative Education 21}, 3 (July 2023), 112--122.
\newblock \url{https://onlinelibrary.wiley.com/doi/10.1111/dsji.12282}.
\newblock \href {https://doi.org/10.1111/dsji.12282} {\path{doi:10.1111/dsji.12282}}.

\bibitem[{\v C}P23]{CernyPotancokInformationLiteracyInternational2023}
\textsc{{\v C}ern{\'y} J., Potan{\v c}ok M.}:
\newblock Information literacy in international masters students: {{A}} competitive and business intelligence course perspective.
\newblock \emph{Cogent Education 10}, 1 (Dec. 2023), 2161701.
\newblock \url{https://www.tandfonline.com/doi/full/10.1080/2331186X.2022.2161701}.
\newblock \href {https://doi.org/10.1080/2331186X.2022.2161701} {\path{doi:10.1080/2331186X.2022.2161701}}.

\bibitem[CRDBL04]{CoxetalCognitiveProcessingPerspective2004}
\textsc{Cox R., Romero P., Du~Boulay B., Lutz R.}:
\newblock A {{Cognitive Processing Perspective}} on {{Student Programmers}}' `{{Graphicacy}}'.
\newblock In \emph{Diagrammatic {{Representation}} and {{Inference}}}, Goos G., Hartmanis J., Van~Leeuwen J., Blackwell A.~F., Marriott K., Shimojima A., (Eds.), vol.~2980. Springer Berlin Heidelberg, Berlin, Heidelberg, 2004, pp.~344--346.
\newblock \url{http://link.springer.com/10.1007/978-3-540-25931-2_35}.
\newblock \href {https://doi.org/10.1007/978-3-540-25931-2_35} {\path{doi:10.1007/978-3-540-25931-2_35}}.

\bibitem[CS98]{CarpenterShahModelPerceptualConceptual1998}
\textsc{Carpenter P.~A., Shah P.}:
\newblock A model of the perceptual and conceptual processes in graph comprehension.
\newblock \emph{Journal of Experimental Psychology: Applied 4}, 2 (June 1998), 75--100.
\newblock \url{https://doi.apa.org/doi/10.1037/1076-898X.4.2.75}.
\newblock \href {https://doi.org/10.1037/1076-898X.4.2.75} {\path{doi:10.1037/1076-898X.4.2.75}}.

\bibitem[Cur87]{curcioComprehensionMathematicalRelationships1987}
\textsc{Curcio F.~R.}:
\newblock Comprehension of {{Mathematical Relationships Expressed}} in {{Graphs}}.
\newblock \emph{Journal for Research in Mathematics Education 18}, 5 (1987), 382--393.
\newblock \url{https://www.jstor.org/stable/749086}.
\newblock \href {http://arxiv.org/abs/749086} {\path{arXiv:749086}}, \href {https://doi.org/10.2307/749086} {\path{doi:10.2307/749086}}.

\bibitem[DBALI18]{DanieletalDefinitionRepresentationalCompetence2018}
\textsc{Daniel K.~L., Bucklin C.~J., Austin~Leone E., Idema J.}:
\newblock Towards a {{Definition}} of {{Representational Competence}}.
\newblock In \emph{Towards a {{Framework}} for {{Representational Competence}} in {{Science Education}}}, Daniel K.~L., (Ed.), vol.~11. Springer International Publishing, Cham, 2018, pp.~3--11.
\newblock \url{http://link.springer.com/10.1007/978-3-319-89945-9_1}.
\newblock \href {https://doi.org/10.1007/978-3-319-89945-9_1} {\path{doi:10.1007/978-3-319-89945-9_1}}.

\bibitem[D'I22]{DIgnazioCreativeDataLiteracy2022}
\textsc{D'Ignazio C.}:
\newblock Creative data literacy: {{Bridging}} the gap between the data-haves and data-have nots.
\newblock \emph{Information Design Journal} (July 2022), 6--18.
\newblock \url{http://www.jbe-platform.com/content/journals/10.1075/idj.23.1.03dig}.
\newblock \href {https://doi.org/10.1075/idj.23.1.03dig} {\path{doi:10.1075/idj.23.1.03dig}}.

\bibitem[DM19]{DobrinMoreyMediatingNatureRole2019}
\textsc{Dobrin S.~I., Morey S.} (Eds.):
\newblock \emph{Mediating {{Nature}}: {{The Role}} of {{Technology}} in {{Ecological Literacy}}}, 1~ed.
\newblock Routledge, Oct. 2019.
\newblock \url{https://www.taylorfrancis.com/books/9780429678172}.
\newblock \href {https://doi.org/10.4324/9780429399121} {\path{doi:10.4324/9780429399121}}.

\bibitem[DYO{\etalchar{*}}20]{DurandetalGraphLiteracyMatters2020}
\textsc{Durand M.-A., Yen R.~W., O'Malley J., Elwyn G., Mancini J.}:
\newblock Graph literacy matters: {{Examining}} the association between graph literacy, health literacy, and numeracy in a {{Medicaid}} eligible population.
\newblock \emph{PLOS ONE 15}, 11 (Nov. 2020), e0241844.
\newblock \url{https://dx.plos.org/10.1371/journal.pone.0241844}.
\newblock \href {https://doi.org/10.1371/journal.pone.0241844} {\path{doi:10.1371/journal.pone.0241844}}.

\bibitem[EHDB14]{EllweinetalUsingRichContext2014}
\textsc{Ellwein A.~L., Hartley L.~M., Donovan S., Billick I.}:
\newblock Using {{Rich Context}} and {{Data Exploration}} to {{Improve Engagement}} with {{Climate Data}} and {{Data Literacy}}: {{Bringing}} a {{Field Station}} into the {{College Classroom}}.
\newblock \emph{Journal of Geoscience Education 62}, 4 (Nov. 2014), 578--586.
\newblock \url{https://www.tandfonline.com/doi/full/10.5408/13-034}.
\newblock \href {https://doi.org/10.5408/13-034} {\path{doi:10.5408/13-034}}.

\bibitem[FDL20]{FiratDenisovaLarameeTreemapLiteracyClassroomBased2020}
\textsc{Firat E.~E., Denisova A., Laramee R.~S.}:
\newblock Treemap {{Literacy}}: {{A Classroom-Based Investigation}}.
\newblock \emph{Eurographics 2020 - Education Papers} (2020), 10 pages.
\newblock \url{https://diglib.eg.org/handle/10.2312/eged20201032}.
\newblock \href {https://doi.org/10.2312/EGED.20201032} {\path{doi:10.2312/EGED.20201032}}.

\bibitem[FDWL22]{FiratetalPLiteStudyParallel2022}
\textsc{Firat E.~E., Denisova A., Wilson M.~L., Laramee R.~S.}:
\newblock P-{{Lite}}: {{A}} study of parallel coordinate plot literacy.
\newblock \emph{Visual Informatics 6}, 3 (Sept. 2022), 81--99.
\newblock \url{https://linkinghub.elsevier.com/retrieve/pii/S2468502X22000377}.
\newblock \href {https://doi.org/10.1016/j.visinf.2022.05.002} {\path{doi:10.1016/j.visinf.2022.05.002}}.

\bibitem[FJJ{\etalchar{*}}24]{FuchsetalTreEducationVisualEducation2024}
\textsc{Fuchs J., J{\"a}ckl B., J{\"u}ttler M., Keim D.~A., Sevastjanova R.}:
\newblock {{TreEducation}}: {{A Visual Education Platform}} for {{Teaching Treemap Layout Algorithms}}.
\newblock \emph{IEEE Transactions on Visualization and Computer Graphics} (2024), 1--16.
\newblock \url{https://ieeexplore.ieee.org/document/10508116/}.
\newblock \href {https://doi.org/10.1109/TVCG.2024.3393012} {\path{doi:10.1109/TVCG.2024.3393012}}.

\bibitem[FJL22a]{FiratJoshiLarameeInteractiveVisualizationLiteracy2022}
\textsc{Firat E.~E., Joshi A., Laramee R.~S.}:
\newblock Interactive visualization literacy: {{The}} state-of-the-art.
\newblock \emph{Information Visualization 21}, 3 (July 2022), 285--310.
\newblock \url{https://journals.sagepub.com/doi/10.1177/14738716221081831}.
\newblock \href {https://doi.org/10.1177/14738716221081831} {\path{doi:10.1177/14738716221081831}}.

\bibitem[FJL22b]{FiratJoshiLarameeVisLitEVisualizationLiteracy2022}
\textsc{Firat E.~E., Joshi A., Laramee R.~S.}:
\newblock {{VisLitE}}: {{Visualization Literacy}} and {{Evaluation}}.
\newblock \emph{IEEE Computer Graphics and Applications 42}, 3 (May 2022), 99--107.
\newblock \url{https://ieeexplore.ieee.org/document/9790006/}.
\newblock \href {https://doi.org/10.1109/MCG.2022.3161767} {\path{doi:10.1109/MCG.2022.3161767}}.

\bibitem[FLS{\etalchar{*}}23]{FiratetalConstructivismbasedApproachTreemap2023}
\textsc{Firat E.~E., Lang C., Srinivas B., Peng I., Laramee R.~S., Joshi A.}:
\newblock A {{Constructivism-based Approach}} to {{Treemap Literacy}} in the {{Classroom}}.
\newblock \emph{Eurographics 2023 - Education Papers} (2023), 9--16.
\newblock \url{https://diglib.eg.org/handle/10.2312/eged20231016}.
\newblock \href {https://doi.org/10.2312/EGED.20231016} {\path{doi:10.2312/EGED.20231016}}.

\bibitem[FPS{\etalchar{*}}21]{FranconerietalScienceVisualData2021}
\textsc{Franconeri S.~L., Padilla L.~M., Shah P., Zacks J.~M., Hullman J.}:
\newblock The {{Science}} of {{Visual Data Communication}}: {{What Works}}.
\newblock \emph{Psychological Science in the Public Interest 22}, 3 (Dec. 2021), 110--161.
\newblock \url{https://journals.sagepub.com/doi/10.1177/15291006211051956}.
\newblock \href {https://doi.org/10.1177/15291006211051956} {\path{doi:10.1177/15291006211051956}}.

\bibitem[FRF08]{FaussetRogersFiskUnderstandingRequiredResources2008}
\textsc{Fausset C.~B., Rogers W.~A., Fisk A.~D.}:
\newblock Understanding the {{Required Resources}} in {{Line Graph Comprehension}}.
\newblock \emph{Proceedings of the Human Factors and Ergonomics Society Annual Meeting 52}, 22 (Sept. 2008), 1830--1834.
\newblock \url{https://journals.sagepub.com/doi/10.1177/154193120805202210}.
\newblock \href {https://doi.org/10.1177/154193120805202210} {\path{doi:10.1177/154193120805202210}}.

\bibitem[FS02]{FreedmanShahModelKnowledgeBasedGraph2002}
\textsc{Freedman E.~G., Shah P.}:
\newblock Toward a {{Model}} of {{Knowledge-Based Graph Comprehension}}.
\newblock In \emph{Diagrammatic {{Representation}} and {{Inference}}}, Goos G., Hartmanis J., Van~Leeuwen J., Hegarty M., Meyer B., Narayanan N.~H., (Eds.), vol.~2317. Springer Berlin Heidelberg, Berlin, Heidelberg, 2002, pp.~18--30.
\newblock \url{http://link.springer.com/10.1007/3-540-46037-3_3}.
\newblock \href {https://doi.org/10.1007/3-540-46037-3_3} {\path{doi:10.1007/3-540-46037-3_3}}.

\bibitem[GBMC16]{GrayetalWaysSeeingData2016}
\textsc{Gray J., Bounegru L., Milan S., Ciuccarelli P.}:
\newblock Ways of {{Seeing Data}}: {{Toward}} a {{Critical Literacy}} for {{Data Visualizations}} as {{Research Objects}} and {{Research Devices}}.
\newblock In \emph{Innovative {{Methods}} in {{Media}} and {{Communication Research}}}, Kubitschko S., Kaun A., (Eds.). Springer International Publishing, Cham, 2016, pp.~227--251.
\newblock \url{http://link.springer.com/10.1007/978-3-319-40700-5_12}.
\newblock \href {https://doi.org/10.1007/978-3-319-40700-5_12} {\path{doi:10.1007/978-3-319-40700-5_12}}.

\bibitem[GC13]{Garcia-RetameroCokelyCommunicatingHealthRisks2013}
\textsc{{Garcia-Retamero} R., Cokely E.~T.}:
\newblock Communicating {{Health Risks With Visual Aids}}.
\newblock \emph{Current Directions in Psychological Science 22}, 5 (Oct. 2013), 392--399.
\newblock \url{https://doi.org/10.1177/0963721413491570}.
\newblock \href {https://doi.org/10.1177/0963721413491570} {\path{doi:10.1177/0963721413491570}}.

\bibitem[GC17]{Garcia-RetameroCokelyDesigningVisualAids2017}
\textsc{{Garcia-Retamero} R., Cokely E.~T.}:
\newblock Designing {{Visual Aids That Promote Risk Literacy}}: {{A Systematic Review}} of {{Health Research}} and {{Evidence-Based Design Heuristics}}.
\newblock \emph{Human Factors: The Journal of the Human Factors and Ergonomics Society 59}, 4 (June 2017), 582--627.
\newblock \url{https://journals.sagepub.com/doi/10.1177/0018720817690634}.
\newblock \href {https://doi.org/10.1177/0018720817690634} {\path{doi:10.1177/0018720817690634}}.

\bibitem[GCGJ16]{Garcia-RetameroetalMeasuringGraphLiteracy2016}
\textsc{{Garcia-Retamero} R., Cokely E.~T., Ghazal S., Joeris A.}:
\newblock Measuring {{Graph Literacy}} without a {{Test}}: {{A Brief Subjective Assessment}}.
\newblock \emph{Medical Decision Making 36}, 7 (Oct. 2016), 854--867.
\newblock \url{https://journals.sagepub.com/doi/10.1177/0272989X16655334}.
\newblock \href {https://doi.org/10.1177/0272989X16655334} {\path{doi:10.1177/0272989X16655334}}.

\bibitem[GCK23]{GeCuiKayCALVICriticalThinking2023}
\textsc{Ge L.~W., Cui Y., Kay M.}:
\newblock {{CALVI}}: {{Critical Thinking Assessment}} for {{Literacy}} in {{Visualizations}}.
\newblock In \emph{Proceedings of the 2023 {{CHI Conference}} on {{Human Factors}} in {{Computing Systems}}} (Hamburg Germany, Apr. 2023), ACM, pp.~1--18.
\newblock \url{https://dl.acm.org/doi/10.1145/3544548.3581406}.
\newblock \href {https://doi.org/10.1145/3544548.3581406} {\path{doi:10.1145/3544548.3581406}}.

\bibitem[GCK25]{GeCuiKayAVECAssessmentVisual2025a}
\textsc{Ge L.~W., Cui Y., Kay M.}:
\newblock {{AVEC}}: {{An Assessment}} of {{Visual Encoding Ability}} in {{Visualization Construction}}.
\newblock In \emph{Proceedings of the 2025 {{CHI Conference}} on {{Human Factors}} in {{Computing Systems}}} (New York, NY, USA, Apr. 2025), {{CHI}} '25, Association for Computing Machinery, pp.~1--16.
\newblock \url{https://dl.acm.org/doi/10.1145/3706598.3713364}.
\newblock \href {https://doi.org/10.1145/3706598.3713364} {\path{doi:10.1145/3706598.3713364}}.

\bibitem[GG11]{GalesicGarcia-RetameroGraphLiteracyCrossCultural2011}
\textsc{Galesic M., {Garcia-Retamero} R.}:
\newblock Graph {{Literacy}}: {{A Cross-Cultural Comparison}}.
\newblock \emph{Medical Decision Making 31}, 3 (May 2011), 444--457.
\newblock \url{https://journals.sagepub.com/doi/10.1177/0272989X10373805}.
\newblock \href {https://doi.org/10.1177/0272989X10373805} {\path{doi:10.1177/0272989X10373805}}.

\bibitem[GHC{\etalchar{*}}24]{GeetalMoreComprehensiveUnderstanding2024}
\textsc{Ge L.~W., Hedayati M., Cui Y., Ding Y., Bonilla K., Joshi A., Ottley A., Bach B., Kwon B.~C., Rapp D.~N., Peck E., Padilla L.~M., Correll M., Borkin M.~A., Harrison L., Kay M.}:
\newblock Toward a {{More Comprehensive Understanding}} of {{Visualization Literacy}}.
\newblock In \emph{Extended {{Abstracts}} of the {{CHI Conference}} on {{Human Factors}} in {{Computing Systems}}} (Honolulu HI USA, May 2024), ACM, pp.~1--7.
\newblock \url{https://dl.acm.org/doi/10.1145/3613905.3636289}.
\newblock \href {https://doi.org/10.1145/3613905.3636289} {\path{doi:10.1145/3613905.3636289}}.

\bibitem[Gil13]{GilbertRepresentationsModels2013}
\textsc{Gilbert J.~K.}:
\newblock Representations and {{Models}}.
\newblock In \emph{Constructing {{Representations}} to {{Learn}} in {{Science}}}, Tytler R., Prain V., Hubber P., Waldrip B., (Eds.). SensePublishers, Rotterdam, 2013, pp.~193--198.
\newblock \url{http://link.springer.com/10.1007/978-94-6209-203-7_12}.
\newblock \href {https://doi.org/10.1007/978-94-6209-203-7_12} {\path{doi:10.1007/978-94-6209-203-7_12}}.

\bibitem[GLS11]{GegenfurtnerLehtinenSaljoExpertiseDifferencesComprehension2011}
\textsc{Gegenfurtner A., Lehtinen E., S{\"a}lj{\"o} R.}:
\newblock Expertise {{Differences}} in the {{Comprehension}} of {{Visualizations}}: A {{Meta-Analysis}} of {{Eye-Tracking Research}} in {{Professional Domains}}.
\newblock \emph{Educational Psychology Review 23}, 4 (Dec. 2011), 523--552.
\newblock \url{https://doi.org/10.1007/s10648-011-9174-7}.
\newblock \href {https://doi.org/10.1007/s10648-011-9174-7} {\path{doi:10.1007/s10648-011-9174-7}}.

\bibitem[GR18]{GrillenbergerRomeikeDevelopingTheoreticallyFounded2018}
\textsc{Grillenberger A., Romeike R.}:
\newblock Developing a theoretically founded data literacy competency model.
\newblock In \emph{Proceedings of the 13th {{Workshop}} in {{Primary}} and {{Secondary Computing Education}}} (Potsdam Germany, Oct. 2018), ACM, pp.~1--10.
\newblock \url{https://dl.acm.org/doi/10.1145/3265757.3265766}.
\newblock \href {https://doi.org/10.1145/3265757.3265766} {\path{doi:10.1145/3265757.3265766}}.

\bibitem[GTS10]{GrammelToryStoreyHowInformationVisualization2010}
\textsc{Grammel L., Tory M., Storey M.}:
\newblock How {{Information Visualization Novices Construct Visualizations}}.
\newblock \emph{IEEE Transactions on Visualization and Computer Graphics 16}, 6 (Nov. 2010), 943--952.
\newblock \url{http://ieeexplore.ieee.org/document/5613431/}.
\newblock \href {https://doi.org/10.1109/TVCG.2010.164} {\path{doi:10.1109/TVCG.2010.164}}.

\bibitem[GWL{\etalchar{*}}19]{GableretalDiagramSafariVisualization2019}
\textsc{G{\"a}bler J., Winkler C., Lengyel N., Aigner W., Stoiber C., Wallner G., Kriglstein S.}:
\newblock Diagram {{Safari}}: {{A Visualization Literacy Game}} for {{Young Children}}.
\newblock In \emph{Extended {{Abstracts}} of the {{Annual Symposium}} on {{Computer-Human Interaction}} in {{Play Companion Extended Abstracts}}} (Barcelona Spain, Oct. 2019), ACM, pp.~389--396.
\newblock \url{https://dl.acm.org/doi/10.1145/3341215.3356283}.
\newblock \href {https://doi.org/10.1145/3341215.3356283} {\path{doi:10.1145/3341215.3356283}}.

\bibitem[HA17]{HeAdarVizItCardsCardBasedToolkit2017}
\textsc{He S., Adar E.}:
\newblock {{VizItCards}}: {{A Card-Based Toolkit}} for {{Infovis Design Education}}.
\newblock \emph{IEEE Transactions on Visualization and Computer Graphics 23}, 1 (Jan. 2017), 561--570.
\newblock \url{https://ieeexplore.ieee.org/document/7539629}.
\newblock \href {https://doi.org/10.1109/TVCG.2016.2599338} {\path{doi:10.1109/TVCG.2016.2599338}}.

\bibitem[HBO18]{HerrmannBrumbyOreszczynWattsYourUsage2018}
\textsc{Herrmann M.~R., Brumby D.~P., Oreszczyn T.}:
\newblock Watts your usage? {{A}} field study of householders' literacy for residential electricity data.
\newblock \emph{Energy Efficiency 11}, 7 (Oct. 2018), 1703--1719.
\newblock \url{http://link.springer.com/10.1007/s12053-017-9555-y}.
\newblock \href {https://doi.org/10.1007/s12053-017-9555-y} {\path{doi:10.1007/s12053-017-9555-y}}.

\bibitem[HCS20]{HopkinsCorrellSatyanarayanVisuaLintSketchySitu2020}
\textsc{Hopkins A.~K., Correll M., Satyanarayan A.}:
\newblock {{VisuaLint}}: {{Sketchy In Situ Annotations}} of {{Chart Construction Errors}}.
\newblock \emph{Computer Graphics Forum 39}, 3 (June 2020), 219--228.
\newblock \url{https://onlinelibrary.wiley.com/doi/10.1111/cgf.13975}.
\newblock \href {https://doi.org/10.1111/cgf.13975} {\path{doi:10.1111/cgf.13975}}.

\bibitem[HHK24]{HedayatiHuntKayPixelsPracticesReconceptualizing2024}
\textsc{Hedayati M., Hunt A., Kay M.}:
\newblock From pixels to practices: {{Reconceptualizing}} visualization literacy.
\newblock \url{https://osf.io/6mq42}, May 2024.
\newblock \href {https://doi.org/10.31219/osf.io/6mq42} {\path{doi:10.31219/osf.io/6mq42}}.

\bibitem[HK23]{HedayatiKayChooseyourownD3Labs2023}
\textsc{Hedayati M., Kay M.}:
\newblock ``{{Choose-your-own}}'' {{D3}} labs for learning to adapt online code.
\newblock In \emph{2023 {{IEEE VIS Workshop}} on {{Visualization Education}}, {{Literacy}}, and {{Activities}} ({{EduVis}})} (Oct. 2023), pp.~49--54.
\newblock \url{https://ieeexplore.ieee.org/document/10344069}.
\newblock \href {https://doi.org/10.1109/EduVis60792.2023.00014} {\path{doi:10.1109/EduVis60792.2023.00014}}.

\bibitem[HK25]{HedayatiKayWhatUniversityStudents2025}
\textsc{Hedayati M., Kay M.}:
\newblock What {{University Students Learn In Visualization Classes}}.
\newblock \emph{IEEE Transactions on Visualization and Computer Graphics 31}, 1 (Jan. 2025), 1072--1082.
\newblock \url{https://ieeexplore.ieee.org/document/10678815/}.
\newblock \href {https://doi.org/10.1109/TVCG.2024.3456291} {\path{doi:10.1109/TVCG.2024.3456291}}.

\bibitem[HP11]{HuesteggePhilippEffectsSpatialCompatibility2011}
\textsc{Huestegge L., Philipp A.~M.}:
\newblock Effects of spatial compatibility on integration processes in graph comprehension.
\newblock \emph{Attention, Perception, \& Psychophysics 73}, 6 (Aug. 2011), 1903--1915.
\newblock \url{http://link.springer.com/10.3758/s13414-011-0155-1}.
\newblock \href {https://doi.org/10.3758/s13414-011-0155-1} {\path{doi:10.3758/s13414-011-0155-1}}.

\bibitem[HP18]{HuesteggePotzschIntegrationProcessesFrequency2018}
\textsc{Huestegge L., P{\"o}tzsch T.~H.}:
\newblock Integration processes during frequency graph comprehension: {{Performance}} and eye movements while processing tree maps versus pie charts.
\newblock \emph{Applied Cognitive Psychology 32}, 2 (Mar. 2018), 200--216.
\newblock \url{https://onlinelibrary.wiley.com/doi/10.1002/acp.3396}.
\newblock \href {https://doi.org/10.1002/acp.3396} {\path{doi:10.1002/acp.3396}}.

\bibitem[HWT{\etalchar{*}}23]{HeetalEnthusiasticGroundedAvoidant2023}
\textsc{He H.~A., Walny J., Thoma S., Carpendale S., Willett W.}:
\newblock Enthusiastic and {{Grounded}}, {{Avoidant}} and {{Cautious}}: {{Understanding Public Receptivity}} to {{Data}} and {{Visualizations}}.
\newblock \emph{IEEE Transactions on Visualization and Computer Graphics} (2023), 1--11.
\newblock \url{https://ieeexplore.ieee.org/document/10292909/}.
\newblock \href {https://doi.org/10.1109/TVCG.2023.3326917} {\path{doi:10.1109/TVCG.2023.3326917}}.

\bibitem[J{\"a}n20]{JanickeTeachingIntersectionVisualization2020}
\textsc{J{\"a}nicke S.}:
\newblock Teaching on the {{Intersection}} of {{Visualization}} and {{Digital Humanities}}:.
\newblock In \emph{Proceedings of the 15th {{International Joint Conference}} on {{Computer Vision}}, {{Imaging}} and {{Computer Graphics Theory}} and {{Applications}}} (Valletta, Malta, 2020), {SCITEPRESS - Science and Technology Publications}, pp.~100--109.
\newblock \url{http://www.scitepress.org/DigitalLibrary/Link.aspx?doi=10.5220/0008987101000109}.
\newblock \href {https://doi.org/10.5220/0008987101000109} {\path{doi:10.5220/0008987101000109}}.

\bibitem[JGS22]{JungjohannGebhardtScheerUnderstandingImprovingTeachers2022}
\textsc{Jungjohann J., Gebhardt M., Scheer D.}:
\newblock Understanding and improving teachers' graph literacy for data-based decision-making via video intervention.
\newblock \emph{Frontiers in Education 7} (Sept. 2022), 919152.
\newblock \url{https://www.frontiersin.org/articles/10.3389/feduc.2022.919152/full}.
\newblock \href {https://doi.org/10.3389/feduc.2022.919152} {\path{doi:10.3389/feduc.2022.919152}}.

\bibitem[Ken22]{kendrickInclusiveDesign2022}
\textsc{Kendrick A.}:
\newblock Inclusive {{Design}}.
\newblock \url{https://www.nngroup.com/articles/inclusive-design/}, 2022.

\bibitem[Ker13]{KerrenInformationVisualizationCourses2013}
\textsc{Kerren A.}:
\newblock Information {{Visualization Courses}} for {{Students}} with a {{Computer Science Background}}.
\newblock \emph{IEEE Computer Graphics and Applications 33}, 2 (Mar. 2013), 12--15.
\newblock \url{http://ieeexplore.ieee.org/document/6482532/}.
\newblock \href {https://doi.org/10.1109/MCG.2013.27} {\path{doi:10.1109/MCG.2013.27}}.

\bibitem[KHP{\etalchar{*}}23]{KecketalEduVisWorkshopVisualization2023}
\textsc{Keck M., Huron S., Panagiotidou G., Stoiber C., Rajabiyazdi F., Perin C., Roberts J.~C., Bach B.}:
\newblock {{EduVis}}: {{Workshop}} on {{Visualization Education}}, {{Literacy}}, and {{Activities}}.
\newblock \url{https://arxiv.org/abs/2303.10708}, 2023.
\newblock \href {https://doi.org/10.48550/ARXIV.2303.10708} {\path{doi:10.48550/ARXIV.2303.10708}}.

\bibitem[KHTG14]{KorneretalEyeMovementsIndicate2014}
\textsc{K{\"o}rner C., H{\"o}fler M., Tr{\"o}binger B., Gilchrist I.~D.}:
\newblock Eye {{Movements Indicate}} the {{Temporal Organisation}} of {{Information Processing}} in {{Graph Comprehension}}.
\newblock \emph{Applied Cognitive Psychology 28}, 3 (May 2014), 360--373.
\newblock \url{https://onlinelibrary.wiley.com/doi/10.1002/acp.3006}.
\newblock \href {https://doi.org/10.1002/acp.3006} {\path{doi:10.1002/acp.3006}}.

\bibitem[KLCA14]{KimetalInvestigatingGraphComprehension2014}
\textsc{Kim S., Lombardino L.~J., Cowles W., Altmann L.~J.}:
\newblock Investigating graph comprehension in students with dyslexia: {{An}} eye tracking study.
\newblock \emph{Research in Developmental Disabilities 35}, 7 (July 2014), 1609--1622.
\newblock \url{https://linkinghub.elsevier.com/retrieve/pii/S0891422214001425}.
\newblock \href {https://doi.org/10.1016/j.ridd.2014.03.043} {\path{doi:10.1016/j.ridd.2014.03.043}}.

\bibitem[Kna15]{KnaflicStorytellingDataData2015}
\textsc{Knaflic C.}:
\newblock \emph{Storytelling with Data: A Data Visualization Guide for Business Professionals}.
\newblock Wiley, 2015.
\newblock \url{https://books.google.com/books?id=IheRCgAAQBAJ}.

\bibitem[Kon20]{KongExtendedAbstractWhat2020}
\textsc{Kong Y.}:
\newblock Extended {{Abstract}}: {{What}} to {{Expect When You Are Learning Data Visualization}}: {{A Textual Analysis}} of {{Course Syllabi}}.
\newblock In \emph{2020 {{IEEE International Professional Communication Conference}} ({{ProComm}})} (Kennesaw, GA, USA, July 2020), IEEE, pp.~189--190.
\newblock \url{https://ieeexplore.ieee.org/document/9201248/}.
\newblock \href {https://doi.org/10.1109/ProComm48883.2020.00042} {\path{doi:10.1109/ProComm48883.2020.00042}}.

\bibitem[KSB{\etalchar{*}}23]{KejstovaetalConstructPlayEngaging2023}
\textsc{Kejstov{\'a} M., Stoiber C., Boucher M., Kandlhofer M., Kriglstein S., Aigner W.}:
\newblock Construct and {{Play}}: {{Engaging Students}} with {{Visualizations}} through {{Playful Methods}}.
\newblock In \emph{Companion {{Proceedings}} of the {{Annual Symposium}} on {{Computer-Human Interaction}} in {{Play}}} (Stratford ON Canada, Oct. 2023), ACM, pp.~96--101.
\newblock \url{https://dl.acm.org/doi/10.1145/3573382.3616082}.
\newblock \href {https://doi.org/10.1145/3573382.3616082} {\path{doi:10.1145/3573382.3616082}}.

\bibitem[Lan24]{LanReflectionsTeachingData2024}
\textsc{Lan X.}:
\newblock Reflections on {{Teaching Data Visualization}} at the {{Journalism School}}.
\newblock \url{https://arxiv.org/abs/2408.04386}, 2024.
\newblock \href {https://doi.org/10.48550/ARXIV.2408.04386} {\path{doi:10.48550/ARXIV.2408.04386}}.

\bibitem[LCV{\etalchar{*}}16]{LaietalMeasuringGraphComprehension2016}
\textsc{Lai K., Cabrera J., Vitale J.~M., Madhok J., Tinker R., Linn M.~C.}:
\newblock Measuring {{Graph Comprehension}}, {{Critique}}, and {{Construction}} in {{Science}}.
\newblock \emph{Journal of Science Education and Technology 25}, 4 (Aug. 2016), 665--681.
\newblock \url{http://link.springer.com/10.1007/s10956-016-9621-9}.
\newblock \href {https://doi.org/10.1007/s10956-016-9621-9} {\path{doi:10.1007/s10956-016-9621-9}}.

\bibitem[LKK17]{LeeKimKwonVLATDevelopmentVisualization2017}
\textsc{Lee S., Kim S.-H., Kwon B.~C.}:
\newblock {{VLAT}}: {{Development}} of a {{Visualization Literacy Assessment Test}}.
\newblock \emph{IEEE Transactions on Visualization and Computer Graphics 23}, 1 (Jan. 2017), 551--560.
\newblock \url{http://ieeexplore.ieee.org/document/7539634/}.
\newblock \href {https://doi.org/10.1109/TVCG.2016.2598920} {\path{doi:10.1109/TVCG.2016.2598920}}.

\bibitem[LMQ19]{LoMingQuLearningVisTools2019}
\textsc{Lo L. Y.-H., Ming Y., Qu H.}:
\newblock Learning {{Vis Tools}}: {{Teaching Data Visualization Tutorials}}.
\newblock In \emph{2019 {{IEEE Visualization Conference}} ({{VIS}})} (Vancouver, BC, Canada, Oct. 2019), IEEE, pp.~11--15.
\newblock \url{https://ieeexplore.ieee.org/document/8933751/}.
\newblock \href {https://doi.org/10.1109/VISUAL.2019.8933751} {\path{doi:10.1109/VISUAL.2019.8933751}}.

\bibitem[LWY{\etalchar{*}}16]{LopezetalNursesNumeracyGraphical2016}
\textsc{Lopez K.~D., Wilkie D.~J., Yao Y., Sousa V., Febretti A., Stifter J., Johnson A., Keenan G.~M.}:
\newblock Nurses' {{Numeracy}} and {{Graphical Literacy}}: {{Informing Studies}} of {{Clinical Decision Support Interfaces}}.
\newblock \emph{Journal of Nursing Care Quality 31}, 2 (Apr. 2016), 124--130.
\newblock \url{https://journals.lww.com/00001786-201604000-00005}.
\newblock \href {https://doi.org/10.1097/NCQ.0000000000000149} {\path{doi:10.1097/NCQ.0000000000000149}}.

\bibitem[Ma05]{MaTeachingVisualization2005}
\textsc{Ma K.-L.}:
\newblock Teaching visualization.
\newblock \emph{ACM SIGGRAPH Computer Graphics 39}, 1 (Feb. 2005), 4--5.
\newblock \url{https://dl.acm.org/doi/10.1145/1057792.1057798}.
\newblock \href {https://doi.org/10.1145/1057792.1057798} {\path{doi:10.1145/1057792.1057798}}.

\bibitem[McH24]{McHenryUpdatingCodeTeaching2024}
\textsc{McHenry W.}:
\newblock Updating a {{Code}} for {{Teaching Ethical Visualizations}}.
\newblock \emph{Journal of Management Education 48}, 6 (Dec. 2024), 1090--1120.
\newblock \url{https://journals.sagepub.com/doi/10.1177/10525629241288266}.
\newblock \href {https://doi.org/10.1177/10525629241288266} {\path{doi:10.1177/10525629241288266}}.

\bibitem[MGN{\etalchar{*}}24]{Medley-RathetalFiguresChartsTables2024}
\textsc{{Medley-Rath} S., Gillespie M.~D., Novosel N., Combs S., Fearnow D.}:
\newblock Figures and {{Charts}} and {{Tables}}, {{Oh My}}!: {{A Content Analysis}} of {{Textbook Data Visualizations}}.
\newblock \emph{Teaching Sociology 52}, 3 (July 2024), 257--265.
\newblock \url{https://journals.sagepub.com/doi/10.1177/0092055X231214006}.
\newblock \href {https://doi.org/10.1177/0092055X231214006} {\path{doi:10.1177/0092055X231214006}}.

\bibitem[MH18]{MansoorHarrisonDataVisualizationLiteracy2018}
\textsc{Mansoor H., Harrison L.}:
\newblock Data {{Visualization Literacy}} and {{Visualization Biases}}: {{Cases}} for {{Merging Parallel Threads}}.
\newblock In \emph{Cognitive {{Biases}} in {{Visualizations}}}, Ellis G., (Ed.). Springer International Publishing, Cham, 2018, pp.~87--96.
\newblock \url{http://link.springer.com/10.1007/978-3-319-95831-6_7}.
\newblock \href {https://doi.org/10.1007/978-3-319-95831-6_7} {\path{doi:10.1007/978-3-319-95831-6_7}}.

\bibitem[MHS15]{MalteseHarshSvetinaDataVisualizationLiteracy2015}
\textsc{Maltese A.~V., Harsh J.~A., Svetina D.}:
\newblock Data {{Visualization Literacy}}: {{Investigating Data Interpretation Along}} the {{Novice}}--{{Expert Continuum}}.
\newblock \emph{Journal of College Science Teaching 45}, 1 (Sept. 2015), 84--90.
\newblock \url{https://www.tandfonline.com/doi/full/10.2505/4/jcst15_045_01_84}.
\newblock \href {https://doi.org/10.2505/4/jcst15_045_01_84} {\path{doi:10.2505/4/jcst15_045_01_84}}.

\bibitem[MP86]{McKenziePadillaConstructionValidationTest1986}
\textsc{McKenzie D.~L., Padilla M.~J.}:
\newblock The construction and validation of the test of graphing in science (togs).
\newblock \emph{Journal of Research in Science Teaching 23}, 7 (Oct. 1986), 571--579.
\newblock \url{https://onlinelibrary.wiley.com/doi/10.1002/tea.3660230702}.
\newblock \href {https://doi.org/10.1002/tea.3660230702} {\path{doi:10.1002/tea.3660230702}}.

\bibitem[MPGZ22]{MacedoetalBuildingInformationVisualization2022}
\textsc{Macedo M.~P., Paiva R. O.~A., Gasparini I., Zaina L. A.~M.}:
\newblock Building information visualization of e-learning data with {{Vis2Learning}} guidelines.
\newblock \emph{Journal on Interactive Systems 13}, 1 (Jan. 2022), 42--53.
\newblock \url{https://sol.sbc.org.br/journals/index.php/jis/article/view/1967}.
\newblock \href {https://doi.org/10.5753/jis.2022.1967} {\path{doi:10.5753/jis.2022.1967}}.

\bibitem[Nag20]{NagelVisuallyAnalysingUrban2020}
\textsc{Nagel T.}:
\newblock Visually {{Analysing Urban Mobility}}: {{Results}} and {{Insights}} from {{Three Student Research Projects}}.
\newblock \emph{KN - Journal of Cartography and Geographic Information 70}, 1 (Apr. 2020), 11--18.
\newblock \url{http://link.springer.com/10.1007/s42489-020-00040-5}.
\newblock \href {https://doi.org/10.1007/s42489-020-00040-5} {\path{doi:10.1007/s42489-020-00040-5}}.

\bibitem[NJB]{NJBCMakingBetter}
{{NJBC}}: {{Making Better Graphs}}!
\newblock \url{https://www.dataspire.org/njbc-making-better-graphs}.

\bibitem[NP16]{NolanPerrettTeachingLearningData2016}
\textsc{Nolan D., Perrett J.}:
\newblock Teaching and {{Learning Data Visualization}}: {{Ideas}} and {{Assignments}}.
\newblock \emph{The American Statistician 70}, 3 (July 2016), 260--269.
\newblock \url{https://www.tandfonline.com/doi/full/10.1080/00031305.2015.1123651}.
\newblock \href {https://doi.org/10.1080/00031305.2015.1123651} {\path{doi:10.1080/00031305.2015.1123651}}.

\bibitem[NRDR24]{NagarajRajaDhiliphanRajkumarDevelopmentInquirybasedActive2024}
\textsc{Nagaraj P., Raja M., Dhiliphan~Rajkumar T.}:
\newblock Development of {{Inquiry-based Active Learning Pedagogy Approach}} to {{Heightening Learners}}' {{Critical Thinking Skills}} in {{Data Visualization}} for {{Analytics Course}}.
\newblock \emph{Journal of Engineering Education Transformations 37}, IS2 (Jan. 2024), 295--304.
\newblock \url{https://journaleet.in/download-article.php?Article_Unique_Id=JPR2053&Full_Text_Pdf_Download=True}.
\newblock \href {https://doi.org/10.16920/jeet/2024/v37is2/24053} {\path{doi:10.16920/jeet/2024/v37is2/24053}}.

\bibitem[NZM{\etalchar{*}}24]{NobreetalReadingPixelsInvestigating2024}
\textsc{Nobre C., Zhu K., M{\"o}rth E., Pfister H., Beyer J.}:
\newblock Reading {{Between}} the {{Pixels}}: {{Investigating}} the {{Barriers}} to {{Visualization Literacy}}.
\newblock In \emph{Proceedings of the {{CHI Conference}} on {{Human Factors}} in {{Computing Systems}}} (Honolulu HI USA, May 2024), ACM, pp.~1--17.
\newblock \url{https://dl.acm.org/doi/10.1145/3613904.3642760}.
\newblock \href {https://doi.org/10.1145/3613904.3642760} {\path{doi:10.1145/3613904.3642760}}.

\bibitem[{\"O}AK{\etalchar{*}}24]{OneyetalTestingTestObservations2024}
\textsc{{\"O}ney S., Abdelaal M., Kurzhals K., Betz P., Kropp C., Weiskopf D.}:
\newblock Testing the {{Test}}: {{Observations When Assessing Visualization Literacy}} of {{Domain Experts}}.
\newblock In \emph{2024 {{IEEE Evaluation}} and {{Beyond}} - {{Methodological Approaches}} for {{Visualization}} ({{BELIV}})} (St Pete Beach, FL, USA, Oct. 2024), IEEE, pp.~106--114.
\newblock \url{https://ieeexplore.ieee.org/document/10756054/}.
\newblock \href {https://doi.org/10.1109/BELIV64461.2024.00017} {\path{doi:10.1109/BELIV64461.2024.00017}}.

\bibitem[{\"O}EAM{\"O}18]{OzkanEsraArikanMehmetOzkanStudyVisualizationSkills2018}
\textsc{{\"O}zkan A., Esra~Arikan E., Mehmet~{\"O}zkan E.}:
\newblock A {{Study}} on the {{Visualization Skills}} of 6th {{Grade Students}}.
\newblock \emph{Universal Journal of Educational Research 6}, 2 (Feb. 2018), 354--359.
\newblock \url{http://www.hrpub.org/journals/article_info.php?aid=6800}.
\newblock \href {https://doi.org/10.13189/ujer.2018.060219} {\path{doi:10.13189/ujer.2018.060219}}.

\bibitem[OGCM18]{OkanetalBiasingDebiasingHealth2018}
\textsc{Okan Y., {Garcia-Retamero} R., Cokely E.~T., Maldonado A.}:
\newblock Biasing and debiasing health decisions with bar graphs: {{Costs}} and benefits of graph literacy.
\newblock \emph{Quarterly Journal of Experimental Psychology 71}, 12 (Dec. 2018), 2506--2519.
\newblock \url{https://journals.sagepub.com/doi/10.1177/1747021817744546}.
\newblock \href {https://doi.org/10.1177/1747021817744546} {\path{doi:10.1177/1747021817744546}}.

\bibitem[OJGW19]{OkanetalUsingShortGraph2019}
\textsc{Okan Y., Janssen E., Galesic M., Waters E.~A.}:
\newblock Using the {{Short Graph Literacy Scale}} to {{Predict Precursors}} of {{Health Behavior Change}}.
\newblock \emph{Medical Decision Making 39}, 3 (Apr. 2019), 183--195.
\newblock \url{https://journals.sagepub.com/doi/10.1177/0272989X19829728}.
\newblock \href {https://doi.org/10.1177/0272989X19829728} {\path{doi:10.1177/0272989X19829728}}.

\bibitem[Onl]{OnlineCourseAnalyzing}
Online {{Course}} {\textbar} {{Analyzing}}/{{Presenting Data}}/{{Information}} by {{Edward Tufte}}.
\newblock \url{https://www.edwardtufte.com/online-course/}.

\bibitem[PA15]{PeeblesAliExpertInterpretationBar2015}
\textsc{Peebles D., Ali N.}:
\newblock Expert interpretation of bar and line graphs: The role of graphicacy in reducing the effect of graph format.
\newblock \emph{Frontiers in Psychology 6} (Oct. 2015).
\newblock \url{http://journal.frontiersin.org/Article/10.3389/fpsyg.2015.01673/abstract}.
\newblock \href {https://doi.org/10.3389/fpsyg.2015.01673} {\path{doi:10.3389/fpsyg.2015.01673}}.

\bibitem[PAE19]{PeckAyusoEl-EtrDataPersonalAttitudes2019}
\textsc{Peck E.~M., Ayuso S.~E., {El-Etr} O.}:
\newblock Data is {{Personal}}: {{Attitudes}} and {{Perceptions}} of {{Data Visualization}} in {{Rural Pennsylvania}}.
\newblock In \emph{Proceedings of the 2019 {{CHI Conference}} on {{Human Factors}} in {{Computing Systems}}} (Glasgow Scotland Uk, May 2019), ACM, pp.~1--12.
\newblock \url{https://dl.acm.org/doi/10.1145/3290605.3300474}.
\newblock \href {https://doi.org/10.1145/3290605.3300474} {\path{doi:10.1145/3290605.3300474}}.

\bibitem[Pap80]{PapertMindstormsChildrenComputers1980}
\textsc{Papert S.}:
\newblock \emph{Mindstorms: Children, Computers, and Powerful Ideas}.
\newblock Basic Books, Inc., USA, 1980.

\bibitem[PC19]{PedersenCavigliaDataLiteracyCompound2019}
\textsc{Pedersen A.~Y., Caviglia F.}:
\newblock Data {{Literacy}} as a {{Compound Competence}}.
\newblock In \emph{Digital {{Science}}}, Antipova T., Rocha A., (Eds.), vol.~850. Springer International Publishing, Cham, 2019, pp.~166--173.
\newblock \url{http://link.springer.com/10.1007/978-3-030-02351-5_21}.
\newblock \href {https://doi.org/10.1007/978-3-030-02351-5_21} {\path{doi:10.1007/978-3-030-02351-5_21}}.

\bibitem[PFLJ22]{PengetalEvaluatingBloomsTaxonomybased2022}
\textsc{Peng I., Firat E., Laramee R., Joshi A.}:
\newblock Evaluating {{Bloom}}'s {{Taxonomy-based Learning Modules}} for {{Parallel Coordinates Literacy}}.
\newblock \emph{Eurographics 2022 - Education Papers} (2022), 21--29.
\newblock \url{https://diglib.eg.org/handle/10.2312/eged20221042}.
\newblock \href {https://doi.org/10.2312/EGED.20221042} {\path{doi:10.2312/EGED.20221042}}.

\bibitem[PKH21]{PepplerKeuneHanCultivatingDataVisualization2021}
\textsc{Peppler K., Keune A., Han A.}:
\newblock Cultivating data visualization literacy in museums.
\newblock \emph{Information and Learning Sciences 122}, 1/2 (May 2021), 1--16.
\newblock \url{https://www.emerald.com/insight/content/doi/10.1108/ILS-04-2020-0132/full/html}.
\newblock \href {https://doi.org/10.1108/ILS-04-2020-0132} {\path{doi:10.1108/ILS-04-2020-0132}}.

\bibitem[PL15]{PostigoLopez-ManjonGraphicacyBiologyTextbooks2015}
\textsc{Postigo Y., {L{\'o}pez-Manj{\'o}n} A.}:
\newblock Graphicacy in biology textbooks: Analysis of activities with images / {{Alfabetizaci{\'o}n}} gr{\'a}fica en libros de texto de biolog{\'i}a: An{\'a}lisis de las actividades con im{\'a}genes.
\newblock \emph{Infancia y Aprendizaje 38}, 3 (July 2015), 509--541.
\newblock \url{https://journals.sagepub.com/doi/full/10.1080/02103702.2015.1054667}.
\newblock \href {https://doi.org/10.1080/02103702.2015.1054667} {\path{doi:10.1080/02103702.2015.1054667}}.

\bibitem[PNL{\etalchar{*}}22]{PandelievetalSeriousGameTeaching2022}
\textsc{Pandeliev V., Namanloo A.~A., Lyons K., Bliemel M., {Ali-Hassan} H.}:
\newblock A {{Serious Game}} for {{Teaching Data Literacy}}.
\newblock In \emph{2022 {{IEEE Games}}, {{Entertainment}}, {{Media Conference}} ({{GEM}})} (St. Michael, Barbados, Nov. 2022), IEEE, pp.~1--6.
\newblock \url{https://ieeexplore.ieee.org/document/10017613/}.
\newblock \href {https://doi.org/10.1109/GEM56474.2022.10017613} {\path{doi:10.1109/GEM56474.2022.10017613}}.

\bibitem[PO23]{PandeyOttleyMiniVLATShortEffective2023}
\textsc{Pandey S., Ottley A.}:
\newblock Mini-{{VLAT}}: {{A Short}} and {{Effective Measure}} of {{Visualization Literacy}}.
\newblock \emph{Computer Graphics Forum 42}, 3 (June 2023), 1--11.
\newblock \url{https://onlinelibrary.wiley.com/doi/10.1111/cgf.14809}.
\newblock \href {https://doi.org/10.1111/cgf.14809} {\path{doi:10.1111/cgf.14809}}.

\bibitem[POR16]{PhilipOlivares-PasillasRochaBecomingRaciallyLiterate2016}
\textsc{Philip T.~M., {Olivares-Pasillas} M.~C., Rocha J.}:
\newblock Becoming {{Racially Literate About Data}} and {{Data-Literate About Race}}: {{Data Visualizations}} in the {{Classroom}} as a {{Site}} of {{Racial-Ideological Micro-Contestations}}.
\newblock \emph{Cognition and Instruction 34}, 4 (Oct. 2016), 361--388.
\newblock \url{https://www.tandfonline.com/doi/full/10.1080/07370008.2016.1210418}.
\newblock \href {https://doi.org/10.1080/07370008.2016.1210418} {\path{doi:10.1080/07370008.2016.1210418}}.

\bibitem[PREB13]{PeeblesetalInfluenceGraphSchemas2013}
\textsc{Peebles D., {Ramduny-Ellis} D., Ellis G., Bonner J.~V.}:
\newblock The influence of graph schemas on the interpretation of unfamiliar diagrams.
\newblock \url{https://scholar.google.com/scholar?cluster=13205407528402374728&hl=en&oi=scholarr}.

\bibitem[PWC{\etalchar{*}}22]{PanavasetalJuvenileGraphicalPerception2022}
\textsc{Panavas L., Worth A.~E., Crnovrsanin T., Sathyamurthi T., Cordes S., Borkin M.~A., Dunne C.}:
\newblock Juvenile {{Graphical Perception}}: {{A Comparison}} between {{Children}} and {{Adults}}.
\newblock In \emph{{{CHI Conference}} on {{Human Factors}} in {{Computing Systems}}} (New Orleans LA USA, Apr. 2022), ACM, pp.~1--14.
\newblock \url{https://dl.acm.org/doi/10.1145/3491102.3501893}.
\newblock \href {https://doi.org/10.1145/3491102.3501893} {\path{doi:10.1145/3491102.3501893}}.

\bibitem[QWW{\etalchar{*}}24]{QuadrietalYouSeeWhat2024}
\textsc{Quadri G.~J., Wang A.~Z., Wang Z., Adorno J., Rosen P., Szafir D.~A.}:
\newblock Do {{You See What I See}}? {{A Qualitative Study Eliciting High-Level Visualization Comprehension}}.
\newblock In \emph{Proceedings of the {{CHI Conference}} on {{Human Factors}} in {{Computing Systems}}} (Honolulu HI USA, May 2024), ACM, pp.~1--26.
\newblock \url{https://dl.acm.org/doi/10.1145/3613904.3642813}.
\newblock \href {https://doi.org/10.1145/3613904.3642813} {\path{doi:10.1145/3613904.3642813}}.

\bibitem[RAR{\etalchar{*}}19]{RamasubramanianetalFloodRiskLiteracy2019}
\textsc{Ramasubramanian M., Allan J.~N., Retamero R.~G., {Jenkins-Smith} H., Cokely E.~T.}:
\newblock Flood {{Risk Literacy}}: {{Communication}} and {{Implications}} for {{Protective Action}}.
\newblock \emph{Proceedings of the Human Factors and Ergonomics Society Annual Meeting 63}, 1 (Nov. 2019), 1629--1633.
\newblock \url{https://journals.sagepub.com/doi/10.1177/1071181319631504}.
\newblock \href {https://doi.org/10.1177/1071181319631504} {\path{doi:10.1177/1071181319631504}}.

\bibitem[RB83]{RidingBoardmanRelationshipSexLearning1983}
\textsc{Riding R.~J., Boardman D.~J.}:
\newblock The {{Relationship}} between {{Sex}} and {{Learning Style}} and {{Graphicacy}} in 14-year-old {{Children}}.
\newblock \emph{Educational Review 35}, 1 (Jan. 1983), 69--79.
\newblock \url{http://www.tandfonline.com/doi/abs/10.1080/0013191830350108}.
\newblock \href {https://doi.org/10.1080/0013191830350108} {\path{doi:10.1080/0013191830350108}}.

\bibitem[RCP{\etalchar{*}}24]{RungvivatjarusetalTrainingPediatricPhysicians2024}
\textsc{Rungvivatjarus T., Chong A.~Z., Patel A., Khare M., Bialostozky M., Kuelbs C.~L.}:
\newblock Training pediatric physicians and staff to obtain data from the electronic health record.
\newblock \emph{Healthcare 12}, 1 (Mar. 2024), 100733.
\newblock \url{https://linkinghub.elsevier.com/retrieve/pii/S221307642300060X}.
\newblock \href {https://doi.org/10.1016/j.hjdsi.2023.100733} {\path{doi:10.1016/j.hjdsi.2023.100733}}.

\bibitem[RDDY07]{RushmeieretalRevisitingNeedFormal2007}
\textsc{Rushmeier H., Dykes J., Dill J., Yoon P.}:
\newblock Revisiting the {{Need}} for {{Formal Education}} in {{Visualization}}.
\newblock \emph{IEEE Computer Graphics and Applications 27}, 6 (Nov. 2007), 12--16.
\newblock \url{https://ieeexplore.ieee.org/document/4405649/}.
\newblock \href {https://doi.org/10.1109/MCG.2007.156} {\path{doi:10.1109/MCG.2007.156}}.

\bibitem[RTB08]{RatwaniTraftonBoehm-DavisThinkingGraphicallyConnecting2008}
\textsc{Ratwani R.~M., Trafton J.~G., {Boehm-Davis} D.~A.}:
\newblock Thinking graphically: {{Connecting}} vision and cognition during graph comprehension.
\newblock \emph{Journal of Experimental Psychology: Applied 14}, 1 (2008), 36--49.
\newblock \url{https://doi.apa.org/doi/10.1037/1076-898X.14.1.36}.
\newblock \href {https://doi.org/10.1037/1076-898X.14.1.36} {\path{doi:10.1037/1076-898X.14.1.36}}.

\bibitem[RTM24]{ReadingTurchioeMangalHealthLiteracyNumeracy2024}
\textsc{Reading~Turchioe M., Mangal S.}:
\newblock Health literacy, numeracy, graph literacy, and digital literacy: An overview of definitions, evaluation methods, and best practices.
\newblock \emph{European Journal of Cardiovascular Nursing 23}, 4 (May 2024), 423--428.
\newblock \url{https://academic.oup.com/eurjcn/article/23/4/423/7244654}.
\newblock \href {https://doi.org/10.1093/eurjcn/zvad085} {\path{doi:10.1093/eurjcn/zvad085}}.

\bibitem[SA06]{SchonbornAndersonImportanceVisualLiteracy2006}
\textsc{Sch{\"o}nborn K.~J., Anderson T.~R.}:
\newblock The importance of visual literacy in the education of biochemists*.
\newblock \emph{Biochemistry and Molecular Biology Education 34}, 2 (Mar. 2006), 94--102.
\newblock \url{https://iubmb.onlinelibrary.wiley.com/doi/10.1002/bmb.2006.49403402094}.
\newblock \href {https://doi.org/10.1002/bmb.2006.49403402094} {\path{doi:10.1002/bmb.2006.49403402094}}.

\bibitem[SBN19]{StenlidenBodenNissenStudentsProducersInteractive2019}
\textsc{Stenliden L., Bod{\'e}n U., Nissen J.}:
\newblock Students as {{Producers}} of {{Interactive Data Visualizations}}---{{Digitally Skilled}} to {{Make Their Voices Heard}}.
\newblock \emph{Journal of Research on Technology in Education 51}, 2 (Apr. 2019), 101--117.
\newblock \url{https://www.tandfonline.com/doi/full/10.1080/15391523.2018.1564636}.
\newblock \href {https://doi.org/10.1080/15391523.2018.1564636} {\path{doi:10.1080/15391523.2018.1564636}}.

\bibitem[SC14]{StoferCheComparingExpertsNovices2014}
\textsc{Stofer K., Che X.}:
\newblock Comparing {{Experts}} and {{Novices}} on {{Scaffolded Data Visualizations}} using {{Eye-tracking}}.
\newblock \emph{Journal of Eye Movement Research 7}, 5 (Dec. 2014).
\newblock \url{https://bop.unibe.ch/JEMR/article/view/2389}.
\newblock \href {https://doi.org/10.16910/jemr.7.5.2} {\path{doi:10.16910/jemr.7.5.2}}.

\bibitem[SER{\etalchar{*}}23]{StoiberetalDAnoNoLearningEnvironment2023}
\textsc{Stoiber C., Emrich {\v S}., Radkohl S., Goldgruber E., Aigner W.}:
\newblock {{dAn-oNo}}: {{Learning Environment}} for {{Data Journalists Teaching Data Analytics Principles}}.
\newblock In \emph{2023 {{IEEE VIS Workshop}} on {{Visualization Education}}, {{Literacy}}, and {{Activities}} ({{EduVis}})} (Melbourne, Australia, Oct. 2023), IEEE, pp.~41--48.
\newblock \url{https://ieeexplore.ieee.org/document/10344255/}.
\newblock \href {https://doi.org/10.1109/EduVis60792.2023.00013} {\path{doi:10.1109/EduVis60792.2023.00013}}.

\bibitem[SFN{\etalchar{*}}19]{ShahetalAssociationHealthLiteracy2019}
\textsc{Shah A., {Ferri-Guerra} J., Nadeem M.~Y., Salguero D., {Aparicio-Ugarriza} R., Desir M., Ruiz J.~G.}:
\newblock The association of health literacy, numeracy and graph literacy with frailty.
\newblock \emph{Aging Clinical and Experimental Research 31}, 12 (Dec. 2019), 1827--1832.
\newblock \url{http://link.springer.com/10.1007/s40520-019-01182-x}.
\newblock \href {https://doi.org/10.1007/s40520-019-01182-x} {\path{doi:10.1007/s40520-019-01182-x}}.

\bibitem[SGL19]{StrobelGrundLindnerSeductiveDetailsTheir2019}
\textsc{Strobel B., Grund S., Lindner M.~A.}:
\newblock Do seductive details do their damage in the context of graph comprehension? {{Insights}} from eye movements.
\newblock \emph{Applied Cognitive Psychology 33}, 1 (Jan. 2019), 95--108.
\newblock \url{https://onlinelibrary.wiley.com/doi/10.1002/acp.3491}.
\newblock \href {https://doi.org/10.1002/acp.3491} {\path{doi:10.1002/acp.3491}}.

\bibitem[SH02]{ShahHoeffnerNoTitleFound2002}
\textsc{Shah P., Hoeffner J.}:
\newblock [{{No}} title found].
\newblock \emph{Educational Psychology Review 14}, 1 (2002), 47--69.
\newblock \url{http://link.springer.com/10.1023/A:1013180410169}.
\newblock \href {https://doi.org/10.1023/A:1013180410169} {\path{doi:10.1023/A:1013180410169}}.

\bibitem[SHB08]{StewartHunterBestRelationshipGraphComprehension2008}
\textsc{Stewart B.~M., Hunter A.~C., Best L.~A.}:
\newblock The {{Relationship}} between {{Graph Comprehension}} and {{Spatial Imagery}}: {{Support}} for an {{Integrative Theory}} of {{Graph Cognition}}.
\newblock In \emph{Diagrammatic {{Representation}} and {{Inference}}}, Stapleton G., Howse J., Lee J., (Eds.), vol.~5223. Springer Berlin Heidelberg, Berlin, Heidelberg, 2008, pp.~415--418.
\newblock \url{http://link.springer.com/10.1007/978-3-540-87730-1_52}.
\newblock \href {https://doi.org/10.1007/978-3-540-87730-1_52} {\path{doi:10.1007/978-3-540-87730-1_52}}.

\bibitem[Shr18]{ShreinerDataLiteracySocial2018}
\textsc{Shreiner T.~L.}:
\newblock Data {{Literacy}} for {{Social Studies}}: {{Examining}} the {{Role}} of {{Data Visualizations}} in {{K}}--12 {{Textbooks}}.
\newblock \emph{Theory \& Research in Social Education 46}, 2 (Apr. 2018), 194--231.
\newblock \url{https://www.tandfonline.com/doi/full/10.1080/00933104.2017.1400483}.
\newblock \href {https://doi.org/10.1080/00933104.2017.1400483} {\path{doi:10.1080/00933104.2017.1400483}}.

\bibitem[Shr20]{ShreinerDataliterateCitizenryHow2020}
\textsc{Shreiner T.~L.}:
\newblock Data-literate citizenry: How {{US}} state standards address data and data visualizations in social studies.
\newblock \emph{Information and Learning Sciences 121}, 11/12 (Nov. 2020), 909--931.
\newblock \url{https://www.emerald.com/insight/content/doi/10.1108/ILS-03-2020-0054/full/html}.
\newblock \href {https://doi.org/10.1108/ILS-03-2020-0054} {\path{doi:10.1108/ILS-03-2020-0054}}.

\bibitem[SM24]{ShreinerMartellMakingRaceRacism2024}
\textsc{Shreiner T.~L., Martell C.~C.}:
\newblock Making race and racism invisible: A critical race analysis of data visualizations in online curricular materials for teaching history.
\newblock \emph{Race Ethnicity and Education 27}, 7 (Nov. 2024), 989--1009.
\newblock \url{https://www.tandfonline.com/doi/full/10.1080/13613324.2022.2106473}.
\newblock \href {https://doi.org/10.1080/13613324.2022.2106473} {\path{doi:10.1080/13613324.2022.2106473}}.

\bibitem[Smi22]{SmithHowChartsWork2022}
\textsc{Smith A.}:
\newblock \emph{How Charts Work: {{Understand}} and Explain Data with Confidence}.
\newblock Pearson Education, 2022.
\newblock \url{https://books.google.com/books?id=7RLQEAAAQBAJ}.

\bibitem[Sol22]{SolenScopingFutureVisualization2022}
\textsc{Solen M.}:
\newblock Scoping the {{Future}} of {{Visualization Literacy}}: {{A Review}}.
\newblock \url{https://osf.io/eypgm}, Aug. 2022.
\newblock \href {https://doi.org/10.31219/osf.io/eypgm} {\path{doi:10.31219/osf.io/eypgm}}.

\bibitem[SOX{\etalchar{*}}24]{SeoetalDesigningBornAccessibleCourses2024}
\textsc{Seo J.~Y., O'Modhrain S., Xia Y., Kamath S.~S., Lee B., Coughlan J.}:
\newblock Designing {{Born-Accessible Courses}} in {{Data Science}} and {{Visualization}}: {{Challenges}} and {{Opportunities}} of a {{Remote Curriculum Taught}} by {{Blind Instructors}} to {{Blind Students}}.
\newblock \emph{EuroVis 2024 - Education Papers} (2024).
\newblock \url{https://diglib.eg.org/handle/10.2312/eved20241053}.
\newblock \href {https://doi.org/10.2312/EVED.20241053} {\path{doi:10.2312/EVED.20241053}}.

\bibitem[SSSJ21]{SpenceetalIncreasingDataKnowledgeArtistic2021}
\textsc{Spence J., Schachter E., Saleem A., Jia B.}:
\newblock Increasing {{Data-Knowledge Through Artistic Representation}}.
\newblock In \emph{{{HCI International}} 2021 - {{Posters}}}, Stephanidis C., Antona M., Ntoa S., (Eds.), vol.~1419. Springer International Publishing, Cham, 2021, pp.~609--624.
\newblock \url{https://link.springer.com/10.1007/978-3-030-78635-9_78}.
\newblock \href {https://doi.org/10.1007/978-3-030-78635-9_78} {\path{doi:10.1007/978-3-030-78635-9_78}}.

\bibitem[Str22]{StrantzDataBrickolageTeaching2022}
\textsc{Strantz A.}:
\newblock Data as ``{{Brickolage}}'': {{Teaching Material Data Visualization Design}} with {{LEGO}}.
\newblock In \emph{The 40th {{ACM International Conference}} on {{Design}} of {{Communication}}} (Boston MA USA, Oct. 2022), ACM, pp.~130--135.
\newblock \url{https://dl.acm.org/doi/10.1145/3513130.3558990}.
\newblock \href {https://doi.org/10.1145/3513130.3558990} {\path{doi:10.1145/3513130.3558990}}.

\bibitem[SWV{\etalchar{*}}22]{ShafferetalPatientJudgmentsHypertension2022}
\textsc{Shaffer V.~A., Wegier P., Valentine K.~D., Duan S., Canfield S.~M., Belden J.~L., Steege L.~M., Popescu M., Koopman R.~J.}:
\newblock Patient judgments about hypertension control: The role of patient numeracy and graph literacy.
\newblock \emph{Journal of the American Medical Informatics Association 29}, 11 (Oct. 2022), 1829--1837.
\newblock \url{https://academic.oup.com/jamia/article/29/11/1829/6655787}.
\newblock \href {https://doi.org/10.1093/jamia/ocac129} {\path{doi:10.1093/jamia/ocac129}}.

\bibitem[TM10]{TangMojeRelatingMultimodalRepresentations2010}
\textsc{Tang K.-S., Moje E.~B.}:
\newblock Relating {{Multimodal Representations}} to the {{Literacies}} of {{Science}}.
\newblock \emph{Research in Science Education 40}, 1 (Jan. 2010), 81--85.
\newblock \url{https://doi.org/10.1007/s11165-009-9158-5}.
\newblock \href {https://doi.org/10.1007/s11165-009-9158-5} {\path{doi:10.1007/s11165-009-9158-5}}.

\bibitem[TRA21]{TiroRulianaAswiLiteracyDescriptionProbability2021}
\textsc{Tiro M.~A., {Ruliana}, Aswi A.}:
\newblock Literacy {{Description}} of {{Probability}} for the {{Senior Secondary School Students}} in {{Makassar City}}.
\newblock \emph{Journal of Physics: Conference Series 1863}, 1 (Mar. 2021), 012013.
\newblock \url{https://iopscience.iop.org/article/10.1088/1742-6596/1863/1/012013}.
\newblock \href {https://doi.org/10.1088/1742-6596/1863/1/012013} {\path{doi:10.1088/1742-6596/1863/1/012013}}.

\bibitem[TT06]{TrickettTraftonComprehensiveModelGraph2006}
\textsc{Trickett S.~B., Trafton J.~G.}:
\newblock Toward a {{Comprehensive Model}} of {{Graph Comprehension}}: {{Making}} the {{Case}} for {{Spatial Cognition}}.
\newblock In \emph{Diagrammatic {{Representation}} and {{Inference}}}, Hutchison D., Kanade T., Kittler J., Kleinberg J.~M., Mattern F., Mitchell J.~C., Naor M., Nierstrasz O., Pandu~Rangan C., Steffen B., Sudan M., Terzopoulos D., Tygar D., Vardi M.~Y., Weikum G., {Barker-Plummer} D., Cox R., Swoboda N., (Eds.), vol.~4045. Springer Berlin Heidelberg, Berlin, Heidelberg, 2006, pp.~286--300.
\newblock \url{http://link.springer.com/10.1007/11783183_38}.
\newblock \href {https://doi.org/10.1007/11783183_38} {\path{doi:10.1007/11783183_38}}.

\bibitem[VWAVDJ21]{VanWeertetalPreferenceUnderstandingGraphs2021}
\textsc{Van~Weert J.~C., Alblas M.~C., Van~Dijk L., Jansen J.}:
\newblock Preference for and understanding of graphs presenting health risk information. {{The}} role of age, health literacy, numeracy and graph literacy.
\newblock \emph{Patient Education and Counseling 104}, 1 (Jan. 2021), 109--117.
\newblock \url{https://linkinghub.elsevier.com/retrieve/pii/S0738399120303499}.
\newblock \href {https://doi.org/10.1016/j.pec.2020.06.031} {\path{doi:10.1016/j.pec.2020.06.031}}.

\bibitem[Wan22]{WangVisVisualToolkitTeaching2022}
\textsc{Wang C.}:
\newblock {{VisVisual}}: {{A Toolkit}} for {{Teaching}} and {{Learning Data Visualization}}.
\newblock \emph{IEEE Computer Graphics and Applications 42}, 4 (July 2022), 20--26.
\newblock \url{https://ieeexplore.ieee.org/document/9830795/}.
\newblock \href {https://doi.org/10.1109/MCG.2022.3176199} {\path{doi:10.1109/MCG.2022.3176199}}.

\bibitem[WFL24]{WoodFengLazarHealthDataVisualization2024}
\textsc{Wood R., Feng J.~H., Lazar J.}:
\newblock Health {{Data Visualization Literacy Skills}} of {{Young Adults}} with {{Down Syndrome}} and the {{Barriers}} to {{Inference-making}}.
\newblock \emph{ACM Transactions on Accessible Computing 17}, 1 (Mar. 2024), 1--1.
\newblock \url{https://dl.acm.org/doi/10.1145/3648621}.
\newblock \href {https://doi.org/10.1145/3648621} {\path{doi:10.1145/3648621}}.

\bibitem[WHHB18]{WojtonetalBeginBeginningConstructionist2018}
\textsc{Wojton M.~A., Hayde D., Heimlich J.~E., B{\"o}rner K.}:
\newblock Begin at the {{Beginning}}: {{A Constructionist Model}} for {{Interpreting Data Visualizations}}.
\newblock \emph{Curator: The Museum Journal 61}, 4 (Oct. 2018), 559--574.
\newblock \url{https://onlinelibrary.wiley.com/doi/10.1111/cura.12277}.
\newblock \href {https://doi.org/10.1111/cura.12277} {\path{doi:10.1111/cura.12277}}.

\bibitem[Wol15]{WolfeTeachingStudentsFocus2015}
\textsc{Wolfe J.}:
\newblock Teaching {{Students}} to {{Focus}} on the {{Data}} in {{Data Visualization}}.
\newblock \emph{Journal of Business and Technical Communication 29}, 3 (July 2015), 344--359.
\newblock \url{https://journals.sagepub.com/doi/10.1177/1050651915573944}.
\newblock \href {https://doi.org/10.1177/1050651915573944} {\path{doi:10.1177/1050651915573944}}.

\bibitem[Wom14]{WomackDataVisualizationInformation2014}
\textsc{Womack R.}:
\newblock Data {{Visualization}} and {{Information Literacy}}.
\newblock \url{https://scholarship.libraries.rutgers.edu/esploro/outputs/journalArticle/991031550042804646}.
\newblock \href {https://doi.org/10.7282/T3X92CZF} {\path{doi:10.7282/T3X92CZF}}.

\bibitem[ZSS{\etalchar{*}}24]{ZhouetalUsingNetworkVisualizations2024}
\textsc{Zhou M., Steinberg S., Stiso C., Danish J.~A., Craig K.}:
\newblock Using network visualizations to engage elementary students in locally relevant data literacy.
\newblock \emph{Information and Learning Sciences 125}, 3/4 (Apr. 2024), 209--231.
\newblock \url{https://www.emerald.com/insight/content/doi/10.1108/ILS-06-2023-0069/full/html}.
\newblock \href {https://doi.org/10.1108/ILS-06-2023-0069} {\path{doi:10.1108/ILS-06-2023-0069}}.

\end{thebibliography}

\end{document}